\documentclass[manuscript]{acmart}

\usepackage{graphicx}
\usepackage{amsmath}
\usepackage[linesnumbered,lined,boxed,commentsnumbered]{algorithm2e}
\usepackage[caption=false,font=normalsize,labelfont=sf,textfont=sf]{subfig}
\usepackage{url}
\usepackage[dvipsnames]{xcolor}
\usepackage{tabularx}
\usepackage{threeparttable}
\usepackage{listings}
\usepackage{enumitem}

\definecolor{codegreen}{rgb}{0,0.6,0}
\definecolor{codegray}{rgb}{0.5,0.5,0.5}
\definecolor{codepurple}{rgb}{0.58,0,0.82}
\definecolor{backcolour}{rgb}{0.95,0.95,0.92}

\lstdefinestyle{mystyle}{
  backgroundcolor=\color{backcolour},   commentstyle=\color{codegreen},
  keywordstyle=\color{magenta},
  numberstyle=\tiny\color{codegray},
  stringstyle=\color{codepurple},
  basicstyle=\ttfamily\footnotesize,
  breakatwhitespace=false,         
  breaklines=true,                 
  captionpos=b,                    
  keepspaces=true,                 
  numbers=left,                    
  numbersep=5pt,                  
  showspaces=false,                
  showstringspaces=false,
  showtabs=false,                  
  tabsize=2,
  showlines=true
}

\lstdefinestyle{mystyle1}{
  basicstyle=\scriptsize,
  frame = single,
    showlines=true,
}

\begin{document}

\title{QUT: A Unit Testing Framework for Quantum Subroutines}

\author{M. V. Klymenko}
\email{mike.klymenko@data61.csiro.au}
\orcid{0000-0002-4641-8977}
\affiliation{
  \institution{CSIRO's Data61}
  \city{Clayton}
  \state{VIC}
  \country{Australia}
}

\author{T. Hoang} 
\affiliation{
  \institution{CSIRO's Data61}
  \city{Eveleigh}
  \state{NSW}
  \country{Australia}
}
\author{H. T. Nguyen}
\affiliation{
  \institution{School of Computing and Information Systems, The University of Melbourne}
  \city{Parkville}
  \state{VIC}
  \country{Australia}
}

\author{S. A. Wilkinson}
\affiliation{
  \institution{CSIRO's Data61}
  \city{Clayton}
  \state{VIC}
  \country{Australia}
}
\affiliation{
  \institution{School of Science, RMIT University}
  \city{Melbourne}
  \state{VIC}
  \country{Australia}
}

\author{B. Goldozian}
\affiliation{
  \institution{CSIRO's Data61}
  \city{Eveleigh}
  \state{NSW}
  \country{Australia}
}

\author{X. Zhenchang}
\affiliation{
  \institution{CSIRO's Data61}
  \city{Eveleigh}
  \state{NSW}
  \country{Australia}
}

\author{Q. Lu}
\affiliation{
  \institution{CSIRO's Data61}
  \city{Eveleigh}
  \state{NSW}
  \country{Australia}
}
\affiliation{
  \institution{University of New South Wales}
  \city{Sydney}
  \state{NSW}
  \country{Australia}
}
\author{M. Usman}
\affiliation{
  \institution{CSIRO's Data61}
  \city{Eveleigh}
  \state{NSW}
  \country{Australia}
}
\affiliation{
  \institution{School of Physics, The University of Melbourne}
  \city{Melbourne}
  \state{VIC}
  \country{Australia}
}
\author{L. Zhu}
\affiliation{
  \institution{CSIRO's Data61}
  \city{Eveleigh}
  \state{NSW}
  \country{Australia}
}
\affiliation{
  \institution{University of New South Wales}
  \city{Sydney}
  \state{NSW}
  \country{Australia}
}

\renewcommand{\shortauthors}{M.V. Klymenko et al.}

\begin{abstract}
We present the architectural design and prototype implementation of QUT (Quantum Unit Testing), a framework for unit testing of quantum subroutines. The framework prioritizes usability and simplicity, making the complex theoretical foundations of quantum unit testing - including statistical tests and quantum tomography - accessible to users with diverse backgrounds. This is achieved through the implementation of polymorphic probabilistic assertions, whose evaluation methods adapt to the data types of the arguments used in assertion statements, which may vary according to the context-dependent semantics of quantum subroutines. These arguments can be represented as qubit measurement outcomes, density matrices, or Choi matrices. For each type, the architecture integrates a specific testing protocol -- such as quantum process tomography, quantum state tomography, or Pearson's chi-squared test -- while remaining flexible enough to incorporate additional protocols in the future. The framework is built on the Qiskit software stack, providing compatibility with a broad range of quantum hardware backends and simulation platforms. Drawing on the reasoning provided by the denotational semantics of quantum subroutines, this work also highlights the key distinctions between quantum unit testing and its classical counterpart. 
\end{abstract}


\maketitle

\section{Introduction}

Quantum computing has rapidly progressed from experimental demonstrations to programmable platforms accessible via the cloud. In the current Noisy Intermediate-Scale Quantum (NISQ) era, hardware is characterized by limited number of qubits, coherence time and fidelity, which constrain the reliable execution of quantum programs \cite{Preskill2018quantumcomputingin}. Program behavior is not only affected by static hardware properties but also by temporal variability. For example, fluctuations in device parameters such as $T_1$ relaxation times often exhibit temporal autocorrelation, which can degrade circuit fidelity between different calibration cycles \cite{Carroll2022}. Hardware noise and temporal variations in hardware properties, combined with traditional human errors, can cause quantum programs to produce unexpected outputs, thereby requiring their routine verification. This motivates the exploration of testing techniques for quantum software in general. To date, quantum software testing research has explored differential testing \cite{10.1109/ASE51524.2021.9678792}, mutation analysis \cite{9678563, 9844849}, and coverage-driven generation \cite{9678798}. 

At the same time, the modularity of quantum programs is increasing. Languages such as OpenQASM3 \cite{qasm3} support subroutines, enabling reusable building blocks across applications. This trend is a direct result of technological improvements in quantum hardware, which enable the execution of programs involving an increasing number of qubits, as well as longer sequences of operations that can be logically organized into subroutines \cite{10.1145/3656339, klymenko2024architectural}. While modularity accelerates software development by promoting reusability, it also makes the detection and isolation of defects more challenging, motivating the use of unit testing.

The survey conducted in \cite{10.1145/3838727} indicates that most practitioners involved in quantum testing engage in unit testing (88\%), followed by regression testing (54\%) and acceptance testing (54\%). The survey also points out that most of them do not use any dedicated testing tools, instead writing their own tests and dealing primarily with classical software bugs. A key feature that distinguishes quantum subroutines from quantum programs is the ability of subroutines to accept a quantum state as input. In general, a quantum subroutine may  accept arguments in the form of classical parameters, quantum states, or a combination of both, and can map these inputs to outputs that may likewise be classical or quantum, exhibiting either deterministic or non-deterministic behavior \cite{10.1145/3656339, klymenko2025}. Therefore, classical unit testing frameworks cannot be directly applied to quantum subroutines, as the semantics of quantum programs are fundamentally different and significantly more complex. The rich semantics of quantum subroutines give rise to a variety of assertion statements suitable for software verification, with arguments that may include quantum states, measurement outcomes, operator expectation values, probability distribution functions and more \cite{klymenko2025}. As a result, depending on the specific case, unit tests must use different evaluation procedures, referred to here as testing protocols. The work of Huang and Martonosi \cite{Huang2019} was among the first to introduce assertions based on statistical tests, specifically the $\chi^2$-test, applied to classical measurement outcomes, while recognising the diversity of possible assertion statements. Other possible protocols involve the quantum swap test \cite{10463159, 10.1145/3656339} and quantum process and state tomography \cite{chuang1997prescription, PhysRevLett.86.4195, 10.1145/3188745.3188802, Huang2020}. Some of these unit testing methods have recently been benchmarked in \cite{miranskyy2025feasibility}. Among the relevant approaches to quantum unit testing is quantum property-based testing, which extends classical property-based testing techniques to the quantum domain. This approach has been implemented in QuCheck, a property-based testing framework for Qiskit \cite{quckeck}. The authors of \cite{10463159} investigate methods for decomposing quantum circuits into smaller subunits for unit testing. Furthermore, recent advances in quantum unit testing are systematically reviewed and analysed in \cite{10.1145/3656339}, with the aim of establishing general principles for quantum unit testing.

The methodologies adopted in most existing studies can be characterized as extensions of classical unit-testing techniques to the quantum domain, a conclusion also supported by the systematic mapping study conducted in  \cite{garcia2023quantum}. While this approach provides an important foundation for quantum software testing, it could be further strengthened by incorporating advances from quantum information theory. In particular, test generation can benefit from systematic methods that complement empirically motivated \textit{ad hoc} heuristics with formal guarantees. For example, in their discussion of input/output analysis, the authors of \cite{10.1145/3656339} propose practical testing principles involving different approaches to input-state preparation. These principles could be further supported by theoretical results concerning the adequacy of input states for fault detection. Quantum process tomography, for instance, provides a well-established theoretical foundation for selecting input states to reconstruct quantum processes represented by a quantum program. The authors of \cite{10.1145/3656339} also propose a classification of the possible input and output data types of quantum subroutines based on empirical evidence from the literature. This classification provides a useful starting point for characterizing quantum subroutines and could be complemented by a more formal analysis based on the semantics of quantum programming languages \cite{coecke2017picturing, Heunen_book}. Note that, at this early stage of quantum software development, the relatively small number of existing applications and use cases limits the availability of empirical data needed to support reliable research.

In this work, we aim to bridge the gap between quantum information theory and practical quantum unit testing by developing a unit-testing framework that draws on the quantum state space, informationally complete positive operator-valued measures (POVMs), and semantics of quantum subroutines. At the same time, we recognize that implementing an appropriate testing protocol imposes a substantial cognitive load on practitioners testing quantum subroutines, as it requires expertise in both classical and quantum computing, programming languages, and an understanding of quantum program semantics and quantum information theory. Therefore, we propose a software architecture based on polymorphic assertion statements that integrates multiple unit testing protocols and provides a user-friendly interface, eliminating the need for users to understand the underlying implementation details or configure associated parameters.  

We implement the proposed architecture in a prototype system called \texttt{QUT}, a quantum unit testing framework that evaluates data obtained from quantum measurement and provide empirical, protocol-driven unit tests tailored for quantum subroutines. These tests help identify and isolate errors arising from hardware- and software-level defects, including those caused by human error. \texttt{QUT} focuses on practical and executable assertions that are applicable in both simulated and real hardware environments. The incremental contribution of this work lies in defining and implementing polymorphic assertion statements for quantum unit testing. These statements accommodate different semantic instances associated with different data types. The work also demonstrates that contextual information can guide assertion selection to improve bug detection in specific applications. Based on the assertion data types specified by the user, the framework selects the appropriate testing protocol.

Along the route of developing and validating the proposed framework, we address the following research questions:

\begin{enumerate}[label={\textit{RQ.\arabic*}}]
\item How do testing protocols based on statistical tests, state tomography, and process tomography differ in terms of accuracy? 
\item To what extent does hardware noise influence the outcomes of the testing procedures?
\item How can contextual information assist in selecting the most appropriate testing protocol?
\end{enumerate}

The remainder of this paper is organized as follows. Section 2 presents the theoretical foundations of quantum unit testing, covering the denotational semantics of quantum subroutines, contextual information, and polymorphic probabilistic assertions. Section 3 outlines the research methodology, based on the design science research methodology, and defines the evaluation metrics. Section 4 describes the architecture and implementation of QUT as a Python class library. Section 5 illustrates its application through a working example of quantum subroutine testing. Section 6 evaluates the framework in terms of usability and test quality under ideal and noisy simulation conditions. Finally, Section 7 discusses the findings, limitations, and directions for future work.

\section{Theoretical background}

\subsection{Overview of the denotational semantics of quantum subroutines}

Our unit testing approach is based on the notion of local contextual equivalence of quantum subroutines. Contextual equivalence means that two pieces of code are considered equivalent if they can be interchanged in any complete program without altering its observable behavior \cite{10.1007/978-3-662-54434-1_14, 10.1007/978-3-540-71316-6_1}. Local contextual equivalence is a weaker form of (global) contextual equivalence, implying equivalence only within one specific context. This shift from global to local helps to reduce the computational complexity of empirical unit testing for practical applications \cite{klymenko2025}.

\begin{figure*}[t]
    \centering
    \includegraphics[width=0.75\linewidth]{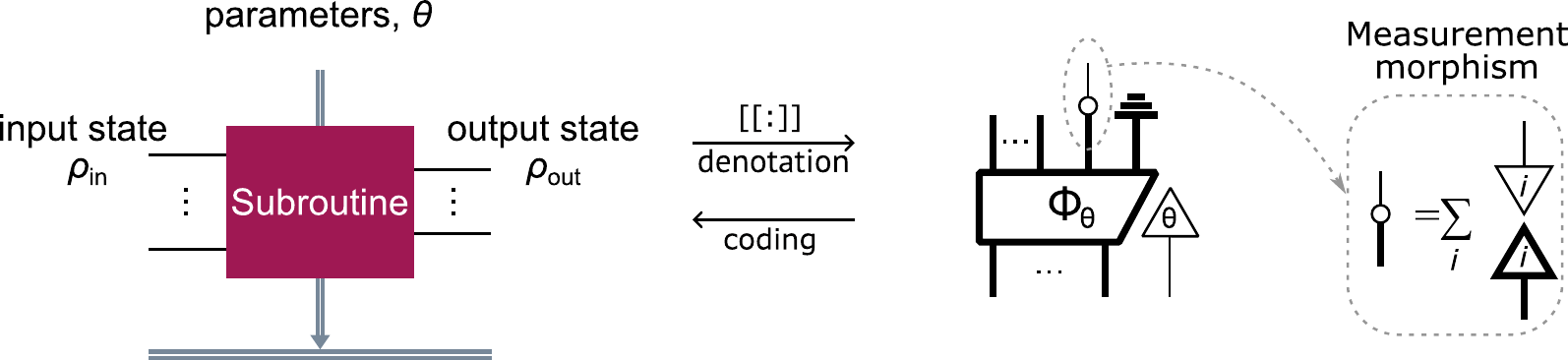}
    \caption{Denotational semantics of a quantum subroutine is defined as a quantum process. The syntax of the subroutine is represented by a circuit diagram. The denotational semantics is expressed using the language of the graphic calculus of monoidal categories for quantum processes \cite{coecke2017picturing}.}
    \label{fig:semantics}
\end{figure*}

Verifying the equivalence of subroutines requires understanding of their semantics. A machine-independent, composable, high-level approach to programming language semantics, known as denotational or mathematical semantics, formalizes the meaning of a program by mapping its syntax to mathematical models \cite{gordon2012denotational}. This semantics is complementary to the operational and axiomatic semantics. The mapping of a syntactic construct to its semantic value is often represented symbolically using the double-bracket operator, where the syntactic expression is placed inside the brackets and its denotational semantics appears outside: $[[syntax]]~=~semantics$. The reverse mapping from semantics to syntax is called coding (see Figure~\ref{fig:semantics}).

Many adaptations of semantics from classical computer science for quantum programs are based on category theory. The categories that are suitable for quantum computing are built by introducing morphisms that capture the distinctive features of quantum information, such as entanglement and the no-cloning theorem \cite{gay2010semantic}. One of the adaptations of category theory for quantum semantics is based on the category \textbf{FdHilb} - the monoidal category of finite-dimensional Hilbert spaces equipped with the symmetric dagger Frobenius structure  \cite{SELINGER2011113, Heunen_book}. Morphisms of this category capture classical and quantum states on the same footing as well as transformations between them including parametrized (classically-controlled) quantum processes, parametrized quantum state preparations and measurements. To incorporate mixed states, this framework must be extended to a new category of completely positive linear maps, \textbf{CP(FdHilb)} that allows discarding of quantum states \cite{coecke2017picturing, Heunen_book}. 

This category is sufficient to serve as the denotational semantics for the quantum subroutines defined as quantum channels in \cite{klymenko2025}. We employ a graphical notation associated with this category to support the reasoning about the unit testing of these subroutines (see Figure~\ref{fig:semantics}). In this notation, the identity morphism associated with a quantum state and its complex conjugate is represented as a thick line, while the identity morphism associated with classical data is shown as a thin line \cite{coecke2017picturing}. Effects and state preparations are depicted as triangles (pointing up and down respectively), and morphisms representing quantum and classical processes are drawn as rectangles. The morphisms associated with discarding quantum information is represented by a ground symbol, analogous to those used in electrical circuit diagrams. In Figure~\ref{fig:semantics}, the core element of the denotational semantics of the subroutine is a parameterized quantum process $\Phi_{\theta}$, where $\theta$ is a vector of parameters. The parameters $\theta$ typically describe rotation angles for quantum unitary gates and can be interpreted as classical inputs to quantum subroutines. This process maps the tensor product of several Hilbert spaces to another Hilbert space, which may have reduced dimensionality if some of the quantum states are discarded or have been subjected to measurements, in accordance with the algorithmic requirements. In addition to the quantum output, the subroutine can also have a classical output, for example, as a result of measurements performed on some qubits, as illustrated in Figure~\ref{fig:semantics}. To indicate that a subroutine with syntax denoted by $\Pi$ has denotational semantics represented by the quantum process $\Phi_{\theta}$, the following notation is used:

\begin{equation}
   \left[\left[ \Pi \right]\right] = \Phi_{\theta}.
\end{equation}

A key requirement of denotational semantics is composability, ensuring that the meaning of complex programs can be derived from the meaning of their constituent parts. Indeed, a quantum program can be represented as a composition of multiple quantum subroutines operating sequentially or in parallel. For instance, it may be expressed as $\Phi_{\theta1}^1 \circ \Phi_{\theta2}^2 \circ \left( I(\rho_a) \otimes \Phi_{\theta3}^3(\rho_0) \right)$, where $I$ denotes the identity channel, $\rho_0$ is the quantum state of a data quantum register, and $\rho_a$ is the quantum register of some ancillary qubits, each $\Phi_{\theta}^j$ represents the quantum channel corresponding to the $j$-th quantum subroutine, and the operation $\circ$ is the composition of two linear maps such that $(A \circ B) (x) = A(B(x))$.

\subsection{Contextual information}

\begin{figure*}
\centering
    \includegraphics[width=0.75\textwidth]{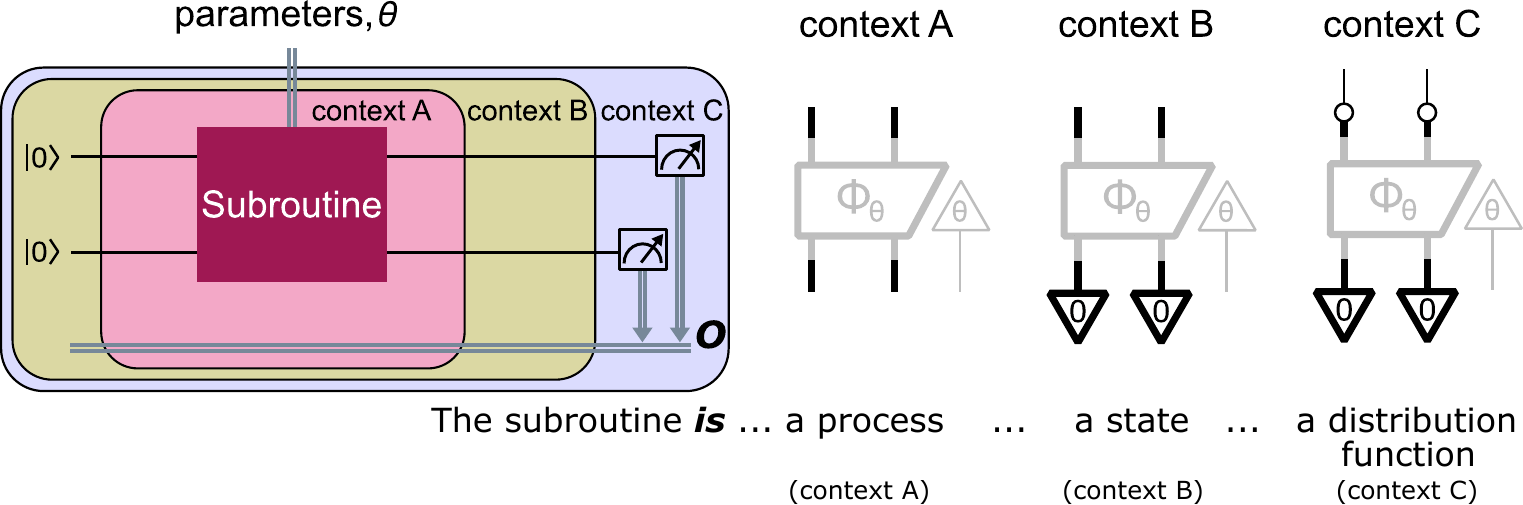}
    \caption{Context-dependent denotational semantics for quantum subroutines for three different contexts: (A) no context information provided (subroutine is used to modify an input quantum state), (B) inputs qubits are initialized to fixed states (subroutine is used as a part of the state preparation process), and (C) same as (B) plus the remaining output qubits subjected to measurement (subroutine is used for sampling probabilistic classical data or evaluating an expectation value). Depending on the environment in which the subroutine is going to be used, the unit test for a quantum subroutine can be significantly simplified imposing modification in the denotational semantics prior to the testing procedure.}
    \label{fig:context}
\end{figure*}

The denotational semantics of quantum programs is characterized by a rich structure of morphisms that describe transformations between various types of object, such as quantum processes, quantum states and vector spaces over binary, real, or complex fields, including both deterministic and random fields. The most informationally complete representation of a quantum subroutine is provided by the notion of a parametrized quantum process (or quantum channel, in the terminology of quantum information theory). Thus, a natural approach to verifying quantum programs is to evaluate the equivalence of formal representations of the quantum channels. These representations include Choi matrices, Pauli transfer matrices, and Kraus operators \cite{Nielsen_Chuang_2010}. Although quantum channels fully capture the logic of quantum subroutines, their use can be impractical, or in many cases even impossible, due to the following reasons: 
\begin{enumerate}
    \item The matrix representations of quantum processes usually scale as $O(2^{4n})$, where $n$ is the number of qubits, making them intractable to store or manipulate for larger systems;
    \item In black-box testing scenarios, the full Choi matrices or other representations of quantum maps are often unavailable or cannot be reconstructed from the specification of quantum subroutine;
    \item An informationally complete description of quantum processes may contain more information than is required for a given application, resulting in unnecessary computational overhead.
\end{enumerate}

The last point suggests that a quantum process - or even a quantum state (density matrix) - may encode information that is excessive for a particular application, with only a portion of that information actually being used. For example, the quantum program in which the subroutine is embedded may act on a quantum state confined to a specific, well-defined subspace of the overall Hilbert space. The knowledge of this information specifies the context in which the subroutine operates. 

The context can be formally described as additional morphisms (quantum processes) applied to the denotational semantics of the subroutine, as illustrated in Figure \ref{fig:context}. These morphisms transform quantum processes associated with quantum subroutines into other entities like quantum states, probability distribution functions of measurement outcomes, random binary numbers, or even binary deterministic numbers. In unit testing, these entities determine the context-dependent semantics of quantum subroutines and serve as arguments in assertion statements. In practice, when unit testing of a specific subroutine is considered, these morphisms represent the denotational semantics of parts of the unit test program that prepare the testing environment and perform post-processing of the subroutine’s output in a way that accounts for context.

\subsection{Polymorphic probabilistic assertions}

As shown above, the context-dependent semantics of quantum subroutines give rise to a variety of outcomes that can serve as arguments in assertion statements within unit tests. A probabilistic assertion is defined as the conditional probability $P(A\vert B)$, where $A$ denotes the assertion statement and $B$ represents the outputs of quantum measurements \cite{klymenko2025}. In this work, we focus on evaluation methods for three types of equality assertion arguments - quantum states, quantum processes, and the distribution functions underlying measurement outcome statistics - all of which can, in principle, be reconstructed from the measurements using appropriate techniques (for instance quantum tomography).

\subsubsection{Equality of quantum states}

Suppose a quantum subroutine is expected to output a quantum state described by the density matrix $\sigma$. Given the data $B$ obtained using a testing protocol, the test outcome can be expressed as the posterior probability of $\sigma$ conditioned on $B$, according to Bayes' theorem:

\begin{equation}
\mbox{Pr}(\sigma \left\vert\right. B) =
\frac{\mbox{Pr}(B \left\vert\right. \sigma)\mbox{Pr}(\sigma)}
{\mbox{Pr}(B)}.
\end{equation}
This expression is not particularly useful in practice because the prior distribution $\Pr(\sigma)$ is generally unknown. A more practical approach is to express the conditional probability as a sum over an auxiliary variable $\rho$:
\begin{equation}
    \mbox{Pr}(\sigma \left\vert\right. B) = \sum_{\rho} \mbox{Pr}(\sigma, \rho \left\vert\right. B) = \sum_{\rho} \mbox{Pr}(\rho \left\vert\right.  B)  \mbox{Pr}(\sigma \left\vert\right. \rho, B) =  \sum_{\rho} \frac{\mbox{Pr}(B \left\vert\right. \rho) \mbox{Pr}(\rho) }{\mbox{Pr}(B)} \mbox{Pr}(\sigma \left\vert\right. \rho, B).
\end{equation}
The sum runs over all states in the discretised state space. Assuming a uniform prior, $\mbox{Pr}(\rho)=\Delta\rho/V_Y$, where $V_Y$ is the volume of the state space and $\Delta\rho$ is the volume element associated with each state, we can convert the sum into an integral by taking the limit $\Delta\rho\to 0$. This yields:
\begin{equation}
    \mbox{Pr}(\sigma \left\vert\right. B) =  \int_{Y} d \rho~\mu_{B} \left(\rho \right) \mbox{Pr}(\sigma \left\vert\right. \rho, B),
    \label{bayes1}
\end{equation}
where
\begin{equation}
\mu_{B} \left( \rho \right)= \frac{1}{V_Y}
\frac{\Pr(B \left\vert\right. \rho)}{\Pr(B)}
\end{equation}
is the normalized likelihood of the observed data $B$ as a function of the candidate state $\rho$, as commonly used in quantum state tomography \cite{Renner_2012}. The integration domain $Y$ encompasses all possible quantum states that can be generated by the unknown quantum subroutine. For a single qubit, $Y$ corresponds to the Bloch ball, with the Bloch sphere representing its boundary of pure states. If the data $B$ has been generated using the quantum tomography, the function has a maximum located at $\rho_{B}$ such that:
\begin{equation}
\rho_{B} = \text{argmax}_{\rho \in Y}~\mu_{B}(\rho ),
\end{equation}
where $\rho_B$ is the density matrix reconstructed from the quantum state tomography. 

The density matrix $\rho$ can be reconstructed with an arbitrary precision provided that the set $\{F_k\}$ constitutes a tomographically complete set of POVM operators and an infinite number of measurements are performed for each operator \cite{PhysRevA.64.052312, PhysRevLett.86.4195}. Since, in practice, only a finite number of measurements can be made, reliable quantum state tomography must be characterized in terms of obtained precision, error bars, and other relevant metrics, the importance of which has been recognized early by several research groups \cite{Smolin_2012, Renner_2012, Renato_2016, Renner_2019}.

If the set of measurements is tomographically complete and a sufficiently large number of samples is collected, the distribution $\mu_{B_n}(\rho)$ can be approximated by the Dirac delta distribution:
\begin{equation}
\mu_{B_n} \left( \rho \right) \approx  \delta\left(\rho - \rho_B \right),
\end{equation}
which leads to a simpler form of Eq.~(\ref{bayes1}):
\begin{equation}
    \mbox{Pr}(\sigma \left\vert\right. B) \approx \mbox{Pr}(\sigma \left\vert\right. \rho_B, B).
\end{equation}
Since $\sigma$ and $B$ are conditionally independent given $\rho_B$, we obtain
\begin{equation}
    \mbox{Pr}(\sigma \left\vert\right. B) \approx \mbox{Pr}(\sigma \left\vert\right. \rho_B).
\end{equation}
Here, $\rho_B$ is the density matrix reconstructed from the quantum state tomography, assuming sufficiently high precision. 

The quantity $\mbox{Pr}(\sigma \left\vert\right. \rho_B)$ can be expressed using the minimum probability of error in distinguishing between two quantum states   
\begin{equation}
\mbox{Pr}(\sigma \left\vert\right. \rho_B) = 2 P_{e,\min} = \left(1 - \big\| \rho_B - \sigma \big\|_{\mathrm{tr}} \right),
\end{equation}

where $\big\| \cdot \big\|_{\mathrm{tr}}$ denotes the trace norm~\cite{Audenaert, tomamichel2015quantum}. The upper bound for the trace distance is provided by one of the Fuchs--van de Graaf inequalities: 
\begin{equation}
1 - \big\| \rho_B - \sigma \big\|_{\mathrm{tr}} \leq F(\rho_B, \sigma),
\end{equation}
where $F(\rho_B, \sigma)$ is the Uhlmann–Jozsa fidelity~\cite{Audenaert, tomamichel2015quantum, UHLMANN1976273,Jozsa01121994}:

\begin{equation}
    \mbox{Pr}(\sigma \left\vert\right. \rho_B) \leq F\left(\rho_{B}, \sigma \right) = \left[ \text{tr} \left( \sqrt{\sqrt{\rho_{B}} \sigma \sqrt{\rho_{B}}} \right) \right]^2.
     \label{eq:fid}
\end{equation}

Thus, the fidelity provides an upper bound on the minimum error probability for distinguishing between two quantum states. Uhlmann and Jozsa also referred to this quantity as transition probability for mixed states. The fidelity has the following properties: $0 \geq  F\left(\rho_B, \sigma \right) \geq 1$; $F\left(\rho_{B}, \sigma \right)=1$ if and only if $\rho_B= \sigma$; $F\left(\rho_{B}, \sigma \right)=0$ if the states are orthogonal. 

\subsubsection{Equality of quantum processes}

Eq.~(\ref{eq:fid}) can also be applied to assess the equivalence of quantum processes, not just quantum states. This is accomplished by replacing the density matrices in Eq.~(\ref{eq:fid}) with normalized Choi matrices, which produces the fidelity of quantum process $P(C \vert C_{B}) \sim F(C_{B}, C)$, where $C$ and $ C_{B}$ are Choi matrices~\cite{PhysRevA.93.042316}. This extension is theoretically supported by channel-state duality, which connects quantum channels and states via the Choi--Jamiolkowski isomorphism~\cite{Wilde_2013}.

\subsubsection{Equality of measurement outcome statistics}

When, instead of assessing the equivalence of the quantum states, we focus on partial information -- such as a specific projection of the state or the expectation value of a particular quantum mechanical operator -- the unit tests can be reduced to verifying the indistinguishability of measurement outcome statistics, $P_{B}$ , and expected statistical model, $P_e$. Unlike the verification of a full quantum state or process, this approach does not require preparing different input states or measuring different state projections.To evaluate the indistinguishability of the probability distribution functions, we adopt the methodology proposed in~\cite{Huang2019}, employing Pearson's $\chi^2$ test as a tool for evaluating probabilistic assertions~\cite{vaughan2013scientific}.  Conceptually, unit testing for measurement outcomes can be viewed as a statistical hypothesis test. The result of the test is characterized by a $p$-value, which quantifies the statistical significance of data~\cite{vaughan2013scientific}. 

A $p$-value, however, cannot be interpreted as a conditional probability that an assertion holds or fails, such as the probability defined in Eq.~(\ref{bayes1}). Nor does it generally provide a lower or upper bound on this probability, as fidelity does (see Eq.~(\ref{eq:fid})). Nevertheless, Ref.~\cite{sellke2001calibration} presents methods for relating (calibrating) $p$-values to error probabilities in hypothesis testing. These methods are based on the statistical distributions of $p$-values across datasets. Specifically, the authors propose identifying a conditioning statistic that quantifies the strength of evidence in the data, either for or against the null hypothesis, and then estimating error probabilities conditional on this statistic. In the present setting, this statistic is defined as $s = \max\{p_0,p_1\}$, where $p_0$ and $p_1$ are the $p$-values obtained by testing $P_B$ against $P_e$ and $P_e$ against $P_B$, respectively. One of the calibration methods described in \cite{sellke2001calibration}, which we adopt in this work, is defined as follows:
\begin{equation}
p^* = \frac{1}{1+\left(s^{-3/4}-1\right)^{1/3}}.
\end{equation}
This calibration transform maps $p$-values to probabilities of Type I error and is characterized by symmetric scaling of Type I and Type II errors. It also introduces a broad uncertainty plateau, where the probability remains close to 0.5 for most  $p$-values, deviating significantly only near the extremes. Furthermore, the transformation preserves the boundary values of the interval, mapping zero to zero and one to one.

\subsubsection{Polymorphism of assertion statements}

Comparing all equality assertions described above, we discover that they share a common feature — in each case, the outcome can be interpreted as the probability that the equality assertion is true. The difference between them lies in how this probability is evaluated. In the framework proposed in this work, we provide a unified specification of equality assertions for all argument types, referred to as polymorphic probabilistic assertions. These assertions adapt their evaluation methods based on the types of arguments supplied. As a result, the evaluation becomes context-dependent, with the context inferred from the argument types.

\subsection{Quantum unit tests and testing protocols}

\begin{figure*}[t]
    \centering
    \includegraphics[width=0.75\linewidth]{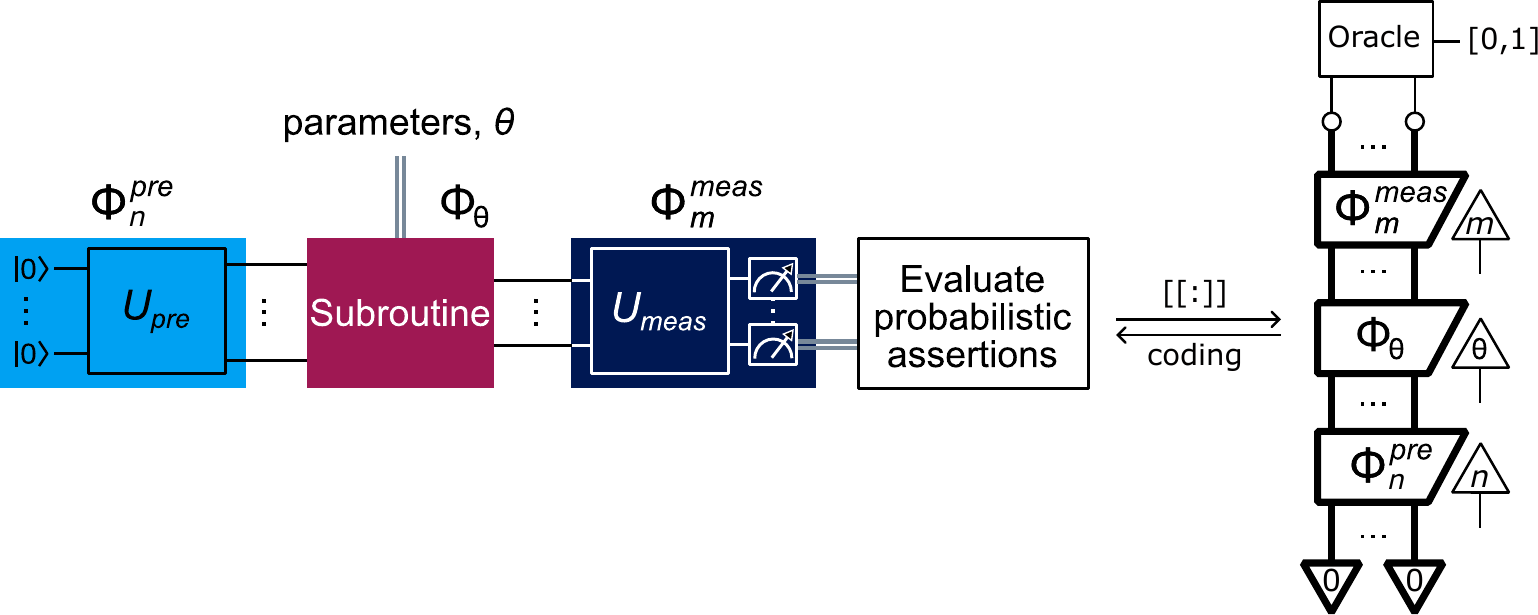}
    \caption{Denotational semantics of a unit test for a quantum subroutine. The explicit forms of the processes $\Phi_n^{pre}$ and $\Phi_m^{meas}$ and the range of the parameters $n$, $\theta$ and $m$ determines the testing protocol. The output of the unit test is a probability of passing the test evaluated by the oracle and expressed as a real number in the closed interval $[0, 1]$.}
    \label{fig:ut}
\end{figure*}

We adopt the arrange-act-assert pattern known from classical unit testing, as has been done in previous studies \cite{klymenko2025, miranskyy2025, 10859187}. During the arrange phase, the testing environment is prepared by constructing the unit test program, which includes the subroutine under test as well as components for preparing input quantum states and post-processing the output. In the act phase, the program is deployed and executed. In the assert phase, the actual output data is analyzed, and the corresponding probabilistic assertions are evaluated. The denotational semantics of each stage corresponds to a quantum process, allowing the entire unit test program to be modeled as a pipeline of quantum channels, as illustrated in Figure~\ref{fig:ut}.

Note that the quantum process corresponding to each stage of unit testing includes many adjustable parameters, such as the rotation angles of quantum gates. Moreover, instead of a single parameter value, a range of parameters can be specified, determining an ensemble of unit test programs. In Figure \ref{fig:ut}, these parameters are represented by vectors $n$, $\theta$ and $m$. The specific values and combinations of these parameters, along with the defined form of the quantum processes $\Phi_n^{\text{pre}}$ and $\Phi_m^{\text{meas}}$ fully determine the semantics of the \textit{quantum testing protocol} in use. For instance, if $n$ is fixed to a single value, and $m$ spans a range of parameters such that $\Phi_m^{\text{meas}}$ forms an informationally complete POVM, the testing protocol corresponds to \textit{quantum state tomography}. If we let $n$ vary such that $\Phi_n^{\text{pre}}$ runs over the complete Pauli basis, we define the testing protocol based on the quantum process tomography. When the vectors $n$ and $m$ are each fixed to a single value, the testing protocol reduces to measuring the output quantum states in the basis specified by $\Phi_{m}^{\text{meas}}$, with the measurement outcome counts serving as the resulting data.  In this case, the testing protocol processes the resulting data using \textit{Pearson's $\chi^2$ statistical test}, which produces a $p$-value as its output, which, in turn, can be calibrated in the probability error $p^*$. Note that this protocol also encompasses an equivalence assertion for the expectation values of quantum operators, since $\Phi_{m}^{\text{meas}}$ can not only rotate the measurement basis but also implement the action of an operator on the output quantum state.

Each testing protocol described above uses a different method to compute the evaluation metric; however, the meaning of the metric is the same in all cases - the conditional probability $P(A \mid B)$ that the assertion statement $A$ is true given the data $B$.

In general case, the development of unit tests and the selection of testing protocols with appropriate reference data involve the following steps:

\begin{enumerate}
    \item Characterize the semantics of the subroutine by identifying the quantum channel it represents, the dimensions of the input and output Hilbert--Schmidt spaces, its input parameters, and any classical outputs.
    
    \item Identify contextual information that may restrict the effective input and output spaces.
    
    \item Select a suitable quantum testing protocol, given the required accuracy, available hardware, and contextual information, which defines the channels $\{\Phi_{\mathrm{pre}}^j\}$ and $\{\Phi_{\mathrm{meas}}^j\}$.
    
    \item Implement a set of unit test programs $\{T_1, T_2, \dots, T_m\}$ by encoding these channels in the chosen programming language.
    
    \item Execute each test on the quantum computer $n$ times and collect the output data $B_n$.
    
    \item Estimate the probabilistic assertion $\Pr(A \mid B_n)$ using a method consistent with the selected protocol.
\end{enumerate}

In this work, we propose the architecture and implementation of a framework called QUT that automates and performs steps 3-6 on behalf of the user.

\section{Methodology}

The methodology of this work is based on the design science research methodology (DSRM) which is a research paradigm focusing on the development and validation of prescriptive knowledge in information science \cite{Peffers01122007}. The DSRM follows a process consisting of the following steps: problem identification and motivation, definition of solution objectives, artifact design and development, demonstration and evaluation, and communication of findings. Note that DSRM is an iterative process that involves repeated cycles of development and evaluation \cite{Peffers01122007}, where each iteration aims to improve the quality of the artifact design. Here, we briefly describe the content of each of these steps.

\subsection{Problem identification and motivation} In classical software engineering, many unit testing frameworks are available. For example, the high-level Python language includes a built-in \texttt{unittest} framework~\cite{python-unittest}, which facilitates writing unit tests with a compact syntax and provides users with assertion methods for different data types. In contrast, to the best of our knowledge, no similar testing framework currently exists for quantum software. This gap results from challenges posed by the complex semantics of quantum subroutines, which in turn give rise to numerous testing protocols.
Each protocol implements trade-offs between the accuracy, coverage, efficiency, and computational complexity of unit tests~\cite{klymenko2025}. A quantum unit testing framework should incorporate a range of these protocols and select the most appropriate one based on the specific context.

\subsection{Definition of solution objectives and evaluation metrics} The key objective of the framework is to facilitate the writing of unit tests for quantum subroutines and to provide methods for evaluating various assertion statements involving different data types (such as quantum states, quantum processes, and deterministic as well as non-deterministic measurement outcomes) used in quantum programming languages. In addition to these main functional requirements, the architectural design of the project is guided by the following technical requirements: the framework shall support multiple quantum hardware systems and backends, and it shall also operate in simulation environments. Another important requirement is that the framework should support a variety of testing protocols and assertion arguments to cover all possible input/output data types. Furthermore, the framework is expected to be expandable by incorporating new testing protocols.

The prioritized quality attributes \cite{bass2003software} relevant to the proposed framework include the following:  \textit{accuracy} - the test outputs should reliably distinguish between correct and incorrect code, and \textit{usability} - the tests shall be written in a concise, economical, and lightweight syntax, with hardware and testing protocol details abstracted away. Usability implies the minimization of cognitive load, assuming that the user has no prior knowledge of quantum information theory.

In classical software engineering, unit testing is typically formulated as a dichotomous problem with a binary outcome. In contrast, quantum unit testing can be formulated as a statistical hypothesis testing problem, where the outcome is usually expressed in terms of probabilities. To quantify the ability of a test with a probabilistic output to discriminate between correct and incorrect code in a dataset containing $N_P$ correct subroutines and $N_N$ mutated subroutines, we use the following metric:
\begin{equation}
    J = \alpha - \beta = \frac{1}{N_P} \sum_{j=1}^{N_P} \alpha_j - \frac{1}{N_P} \sum_{j=1}^{N_N} \beta_j,
    \label{eq:Youden}
\end{equation}
where $\alpha_j$ is predicted probability that the subroutine is correct for the positive example, $\beta_j$ is predicted probability that the subroutine is correct for a negative example. 

This metric can be interpreted as Youden’s J statistic under a frequentist view of probability, where $\alpha$ corresponds to sensitivity, $1-\alpha$ to the false negative rate, $\beta$ to the false positive rate, and $1-\beta$ to specificity. This interpretation is based on the following thought experiment: consider an oracle that outputs a probability (a number between 0 and 1) that the supplied code is correct by effectively performing $N$ experiments, some confirming the correctness of the code and others indicating the presence of mutations. Assume that the correct subroutine is supplied. In this case, $TP$ experiments confirm correctness and $FN$ experiments yield false negatives, so the oracle outputs $TP/(TP + FN)$, with $TP + FN = N$. For negative ground truth, the oracle outputs a probability interpreted as $FP/(TN + FP)$, where $FP$ denotes the number of false positive experiments and $TN$ denotes the number of true negatives, with $TN + FP = N$. Under this interpretation, Eq.~(\ref{eq:Youden}) reduces to the standard definition of Youden’s J statistic:
\begin{equation}
J=\frac{TP}{TP+FN}-\frac{FP}{TN+FP}.
\end{equation}
As more positive or negative cases are included in the dataset, the oracle performs additional experiments and reports average values across the entire dataset.

To evaluate the usability of the framework, as a proxy for developer effort, we measure the complexity of test implementations and associated cognitive load by using the number of lines of code and the number of imported entities (functions and classes) required to implement equivalent tests. Both of these metrics vary depending on the details of a particular implementation, but generally provide a reasonable view of how the code is improved in terms of the number of instructions the user needs to execute to achieve a certain goal and the number of semantic entities he must learn to write the script respectively. It should be noted that the relationship between source code metrics and cognitive load remains an active area of research; see, for example, \cite{ABBADANDALOUSSI2023111619}.

\subsection{Artifact design and development} In this work, the artifacts defined within the DSRM methodology consist of the architecture design for the quantum unit testing framework and its instantiation, called QUT.

\begin{figure*}[t]
    \centering
    \subfloat[]{\includegraphics[width=0.37\linewidth]{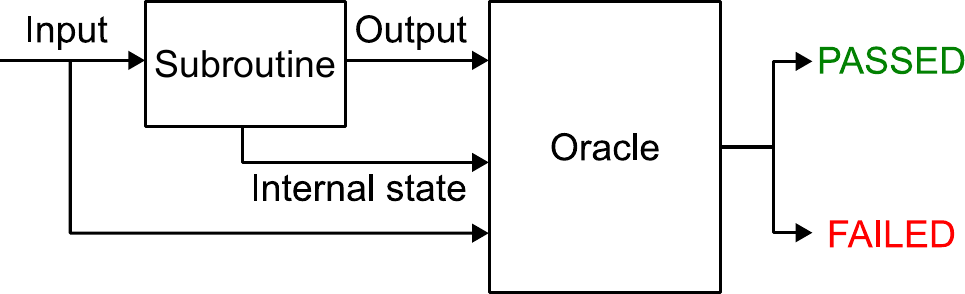}}\\
    \subfloat[]{\includegraphics[width=0.75\linewidth]{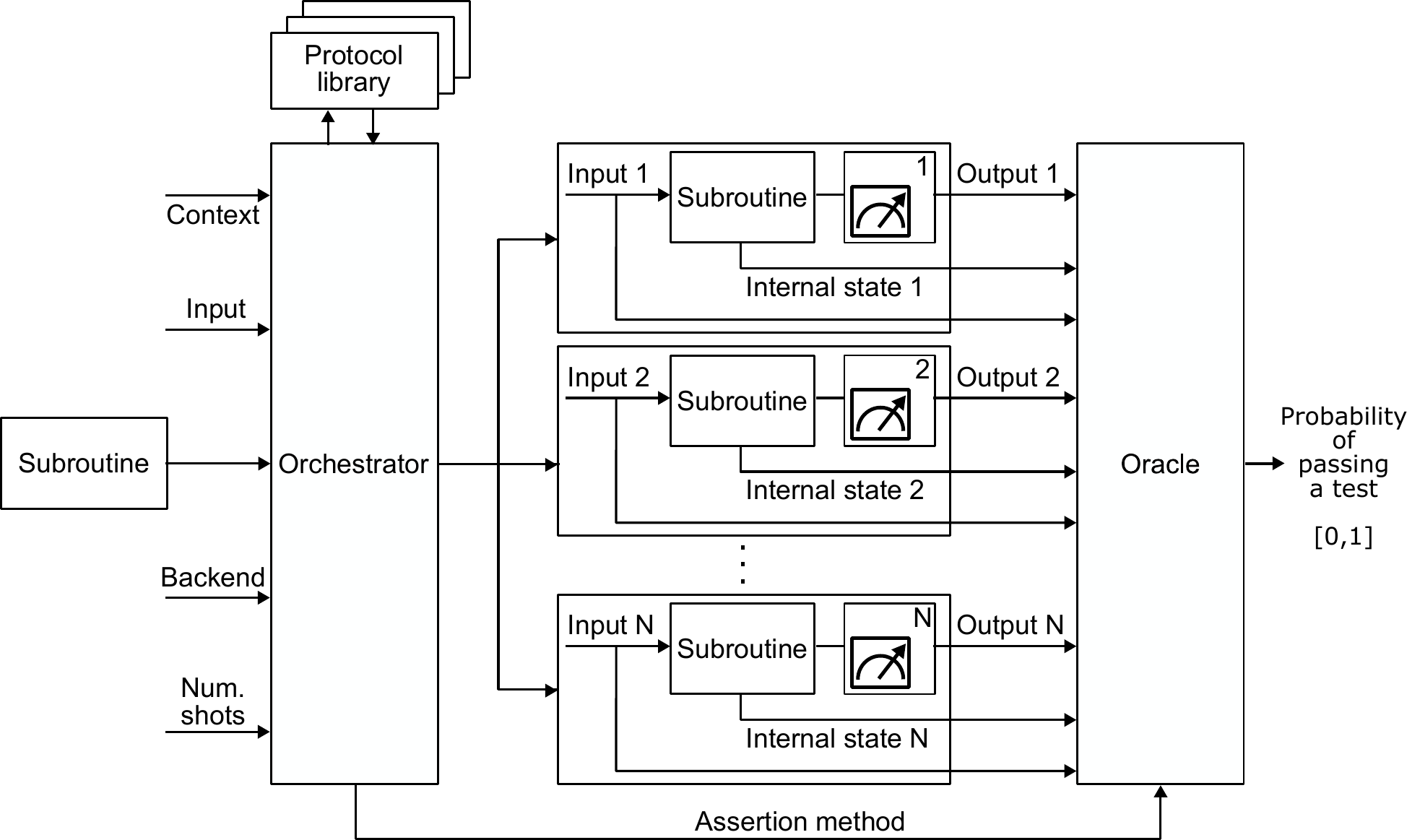}}
    \caption{a) Functional model of a classical unit test; and b) functional model of a quantum unit test with context-aware polymorphic assertions.}
    \label{fig:model}
\end{figure*}

In classical software engineering, a standard functional model for a unit test typically resembles the one shown in Figure~\ref{fig:model}a \cite{bass2003software}. According to this model, a subroutine is executed within a testing computational environment using the input data provided. Upon completion, both the output result and the final state of the subroutine are evaluated using a test oracle. The oracle analyzes this information and determines the test outcome - either passed or failed. 

For quantum software, this model typically does not apply due to the inherent probabilistic nature of quantum information and possibly the presence of quantum correlations. These characteristics make it nearly impossible to evaluate a test assertion based on a single execution of the subroutine. Instead, it is necessary to execute the subroutine multiple times (often referred to as ``shots") to collect statistical data. Additionally, a diverse set of input states must be supplied, and measurements must be performed in different basis sets to reveal quantum correlations or, when quantum state or quantum process tomography is employed \cite{PhysRevLett.86.4195}, to reconstruct the complete density matrix. Preparing many copies of the tested system with different inputs, as well as performing different types of measurements on qubits can be delegated to a program called orchestrator as is illustrated in Figure~\ref{fig:model}b. 

The input states to prepare and types of measurements to apply are specified by a selected testing protocol. This protocol is retrieved from a protocol library and informs the orchestrator about how to configure and execute the testing process accordingly. The testing protocol can be selected either manually by the user or automatically based on the provided input data, test specification, and contextual information regarding how the subroutine is utilized within the broader project. The chosen protocol, along with the input data, test specification, and contextual information, is then provided to the orchestrator. Besides this, given the non-deterministic nature of quantum information, another input to the orchestrator is the number of shots required to generate the statistical output data or, alternatively, the desired accuracy for the test outcomes, from which the appropriate number of shots can be inferred. The number of shots and experiments specified in the protocol instructs the orchestrator to perform 
$N$ runs of the subroutine, using either identical or varying inputs and measurement setups. Additionally, given the diverse characteristics and infrastructures of modern quantum computers, the user must provide the orchestrator with information about the hardware backend - whether it is a simulation environment or an actual quantum computing device. In the latter case, the orchestrator must select the appropriate compiler and scheduler based on the specified backend. It then compiles all generated tests, deploys them onto the hardware, collects the results, and forwards them to the oracle for evaluation. 

The oracle processes all the collected information and evaluates the test assertions according to methods provided by the orchestrator, which in turn obtains them from the specified testing protocol. The results of the test evaluation are presented as output to the user.

Design decisions are based on a set of technical requirements and prioritized quality attributes. In particular, the requirements related to framework extensibility and reusability align closely with the concept of a class library in the object-oriented paradigm. A class library is a collection of predefined classes and interfaces that serve as reusable software components. It is typically provided as part of a programming framework or development platform and can be easily extended through class inheritance. The performance of the framework is based on the notion of context-awareness. To enhance usability, the framework's functionality is embedded into the Python language as an extension and provided as a Python module~\cite{lutz2009learning}.  This choice is driven by Python’s readability, syntactical economy, and dynamic typing, which make it particularly attractive for providing user interfaces to the complex functionality of scientific computations. As a result, it has become the foundation for many scientific-oriented software systems (see, for instance \cite{KLYMENKO2021107676}).

The artifact was developed in three iterative steps, each resulting in improvements according to the evaluation metrics: 

\begin{enumerate}  
    \item \textbf{Initial implementation of unit tests using qiskit-experiments library}: Develop unit tests for different types of assertion arguments using the Qiskit library.  
    
    \item \textbf{Developing unified interface for unit testing and a class library of testing protocols}: Define a unified interface for test protocols and implement them in a dedicated class library. Users write unit tests by creating classes that inherit from a specific protocol implementation relevant to their testing needs.  
    
    \item \textbf{Developing automated protocol selection}: Introduce an orchestrator class that automatically selects the appropriate protocol based on the types of arguments used in the assertions. In this setup, users only need to inherit from a single class when writing unit tests.  
\end{enumerate}  

A detailed description of the framework, whose UML diagram is shown in Figure~\ref{fig:uml}, is provided in Section~\ref{sec:arch}.

\subsection{Demonstration, evaluation, and communication of findings} 

The functionality of the artifact is demonstrated through a working example, with its listing provided in Figure~\ref{fig:rng} in Section~\ref{sec:demo}. This example showcases unit testing of a two-qubit quantum subroutine and its mutated version using three types of equality assertions for quantum process, quantum state, and measurement outcome counts. These demonstrations show how assertion results (e.g., calibrated $p$-values or fidelity scores) vary as a function of the test context, protocol choice, and experimental conditions, validating the framework’s capability to make probabilistic and context-aware test decisions. Notably, the evaluation highlights the benefits of contextual information, as well as the importance of selecting the appropriate testing protocol based on the nature of the subroutine and the available information.

In Section~\ref{sec:demo}, we also evaluate the achievable confidence levels for unit tests in the limit of a large number of shots.

The proposed architectural design has been implemented in a prototype called QUT, an open-source unit testing framework for quantum subroutines available at the link \href{https://github.com/freude/QUT.git}{https://github.com/freude/QUT.git}. The code also contains all the scripts enabling the reproduction of the experiments discussed in this work.

\section{Implementation of QUT as a class library}\label{sec:arch}

\begin{figure*}[t]
    \centering
    \includegraphics[width=\linewidth]{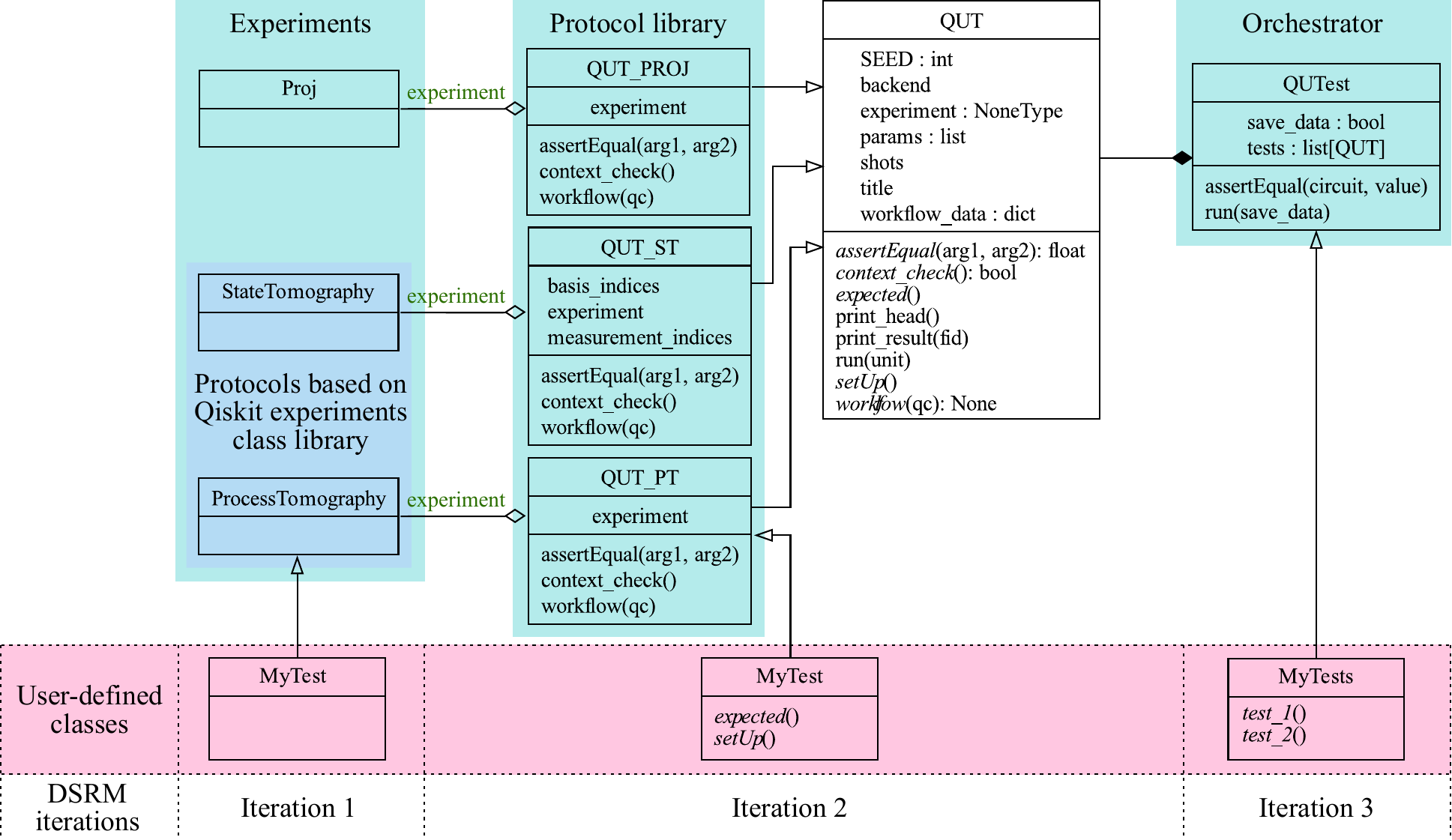}
    \caption{UML diagram of the class library implementing the QUT framework}
    \label{fig:uml}
\end{figure*}

Figure~\ref{fig:uml} illustrates a UML class diagram representing an object-oriented implementation of the framework QUT, whose functional model is shown in Figure~\ref{fig:model}b. The functionality of the orchestrator is implemented in the class \texttt{QUTest}, which serves as the parent class for all user-defined test suits. As shown in the UML diagram, users create test suites by defining a custom class - e.g., \texttt{MyTests} - that inherits from \texttt{QUTest}. Within the class \texttt{MyTests}, the user can create custom methods, each representing an individual test case. The \texttt{QUTest} class provides a predefined method \texttt{run()}, whose purpose is to gather all test cases and execute them within the testing environment. This class also contains a definition of a polymorphic function for evaluating assertions \texttt{assertEqual()}, which the user can use in his test cases. Depending on the type of arguments supplied to \texttt{assertEqual()}, one of the protocols from the protocol library will be invoked.  The class \texttt{QUTest} has a boolean attribute \texttt{save\_data}. If set to \texttt{True}, all intermediate information generated during unit testing - such as density matrices, measurement outcomes, etc. - is stored and made accessible through the \texttt{tests} attribute of the \texttt{QUTest} class instance. The classes \texttt{QUTest}  was developed during the final iteration of the DSRM process.

The interface for the protocols is determined by the class \texttt{QUT}. This is a base abstract class for all the protocols. An abstract class in object-oriented programming serves as a blueprint for other classes, defining a general concept or functionality that can be extended but cannot be directly instantiated. It is a template that specifies which methods must be implemented by its subclasses, ensuring a consistent structure while allowing for specialization. Abstract classes often contain abstract methods (without implementation) and can also include concrete methods (with implementation). Such concrete methods of the base abstract class \texttt{QUT} is the method \texttt{run} implementing the template method pattern \cite{gamma1995design}. The template method remains invariant across all subclasses and implements a sequence of high-level steps, which are delegated to helper methods that are part of the same class and are typically abstract. The template method \texttt{run()} implements general logic of unit testing and defines fixed steps of the arrange–act–assert pattern such as setting up the testing environment, generating and executing tests according to the selected testing protocols, and providing the testing oracle with the required output data and evaluation method. Each of these steps corresponds to abstract methods in the base class and must be defined in the subclasses.

The subclasses of \texttt{QUT} implement specific testing protocols. Each protocol encapsulates information about the preparation of multiple input quantum states and the execution of specific measurements on the output quantum states. In the current version, the protocol library includes three protocol classes: \texttt{QUT\_PROJ}, \texttt{QUT\_ST}, and \texttt{QUT\_PT}, which implement Pearson's $\chi^2$ test, quantum state tomography, and quantum process tomography, respectively. Each of these subclasses must contain a concrete implementation of the equality assertion method \texttt{assertEqual()}, the method \texttt{context\_check()} for determining whether the contextual information matches the protocol, and the specific workflow that defines the testing procedure, implemented in the method \texttt{workflow()}. The classes \texttt{QUT\_PROJ}, \texttt{QUT\_ST}, and \texttt{QUT\_PT} can be used without the orchestrator. In this case, the user creates a test case, represented by a class (\texttt{MyTest} in Figure~\ref{fig:uml}), that inherits from one of these classes. This approach corresponds to the second iteration of the DSRM process and imposes a higher cognitive load, as the user must be familiar with all available protocols in the library and their specifications.

The \texttt{workflow()} method is implemented with the help of the object \texttt{experiment}, which encapsulates the details of the specific interaction with quantum hardware or the simulation environment. For protocols based on quantum tomography, this object is the instance of the \texttt{StateTomography} or \texttt{ProcessTomography} class from the \texttt{qiskit\_experiments} framework \cite{qiskit-experiments}. For Pearson's $\chi^2$ test, this object is the instance of the \texttt{Proj} class, which defines the procedure for performing projective measurements in a specific basis set. We refer to the collection of classes \texttt{Proj}, \texttt{StateTomography}, and \texttt{ProcessTomography} as experiments, since most of them are based on, or share interfaces with, the \texttt{qiskit\_experiments} library for standard quantum characterization, calibration, and verification tasks \cite{qiskit-experiments}. Using these experiments directly to write a unit test program constitutes the first iteration of the DSRM process.

\begin{figure}
    \centering
    \begin{lstlisting}[language=Python]
from qiskit import QuantumCircuit
from qiskit.quantum_info import DensityMatrix, Choi
from qiskit_aer import AerSimulator
from qut import QUTest, detach


@detach
def subroutine_correct(circuit):
    circuit.x(0)      # apply Pauli-X gate to the first qubit
    circuit.h(0)      # apply Hadamard gate to the first qubit
    circuit.cx(0, 1)  # apply CNOT gate
    return circuit


@detach
def subroutine_error(circuit):
    circuit.x(0)      # apply Pauli-X gate to the first qubit
    circuit.h(1)      # apply Hadamard gate to the first qubit
    circuit.cx(0, 1)  # apply CNOT gate
    return circuit


class MyTests(QUTest):
    """Class prepares environment for a quantum unit test."""

    def setUp(self):
        self.qinput = QuantumCircuit(2)
        self.distr = [0.5, 0.0, 0.0, 0.5]
        self.state = DensityMatrix([[0.5, 0.0, 0.0, -0.5],
                                    [0.0, 0.0, 0.0, 0.0],
                                    [0.0, 0.0, 0.0, 0.0],
                                    [-0.5, 0.0, 0.0, 0.5]])
        self.proc = Choi(subroutine_correct(self.qinput))

    def test_1(self):
        self.assertEqual(subroutine_correct(self.qinput), self.distr)
        self.assertEqual(subroutine_correct(self.qinput), self.state)
        self.assertEqual(subroutine_correct(self.qinput), self.proc)

    def test_2(self):
        self.assertEqual(subroutine_error(self.qinput), self.distr)
        self.assertEqual(subroutine_error(self.qinput), self.state)
        self.assertEqual(subroutine_error(self.qinput), self.proc)


# run tests
tests = MyTests(backend=AerSimulator(), shots=3000)
tests.run()
\end{lstlisting}
\begin{lstlisting}[escapeinside={(*@}{@*)}, basicstyle=\small\ttfamily]
(*@\textcolor{ForestGreen}{[PASSED]: with a 0.981 probability of passing.}@*)
(*@\textcolor{ForestGreen}{[PASSED]: with a 0.995 probability of passing.}@*)
(*@\textcolor{ForestGreen}{[PASSED]: with a 0.986 probability of passing.}@*)
(*@\textcolor{red}{[FAILED]: with a 0.000 probability of passing.}@*)
(*@\textcolor{red}{[FAILED]: with a 0.256 probability of passing.}@*)
(*@\textcolor{red}{[FAILED]: with a 0.000 probability of passing.}@*)
\end{lstlisting}
    \caption{Usage example of QUT on a two-qubit subroutine}
    \label{fig:rng}
\end{figure}

The architectural separation between protocol definitions, experiments implementation, and orchestration logic promotes modularity, reusability, and scalability in the development of quantum-classical applications. In this architecture, the protocol library can be easily extended with additional testing protocols, which can be integrated through the interface defined by the \texttt{QUT} class.

Figure~\ref{fig:uml} also illustrates the results of the design iterations within DSRM. In the first iteration, the framework consisted essentially of instantiations of classes from the Qiskit experiment library, which implemented quantum state tomography and quantum process tomography protocols, or user-defined classes that set up experiments using the Qiskit framework. The assertions have been implemented using functions that perform fidelity calculations or statistical tests. In the next iteration, the testing protocols were encapsulated within a single interface with standardized methods. Additionally, a library of classes implementing specific protocols has been created. The user can create tests by instantiating one of these classes using a standardized interface provided for each of them. In the final design iteration, the orchestrator class was introduced to enable context detection and automatic selection of testing protocols. Examples of unit tests corresponding to each of these design iterations can be found at the link \url{https://github.com/freude/QUT/tree/main/examples/DSRM_iterations}.

\section{Usage example for QUT}\label{sec:demo}

The Python script, shown in Figure~\ref{fig:rng} along with its output, illustrates the application of the \texttt{QUT} framework in combination with Qiskit's \texttt{AerSimulator} backend to test a two-qubit subroutine. 

The lines 1-4 of the scripts are library imports. On the line 3, the script imports \texttt{AerSimulator} from the module \texttt{qiskit-aer}, which provides a high-performance quantum backend simulator. The QUT framework provides a Python module called \texttt{qut}, which is imported on line 4.

At lines 8-20, the script defines two subroutines intended as inputs for unit testing: \texttt{subroutine\_correct()}, which serves as a reference implementation and is expected to generate a Bell state when initialized with the input $\vert 00 \rangle$; and \texttt{subroutine\_error()}, which includes an intentionally introduced fault for testing purposes -- specifically, a quantum gate is incorrectly applied to the wrong qubit index, simulating a common implementation error. 

The testing environment along with test cases are defined by the user-defined class \texttt{MyTest}, line 23. The class \texttt{MyTest} is a subclass of the class \texttt{QUTest}, which serves as the test orchestrator and is provided by the module \texttt{qut}. The base class \texttt{QUTest} automatically detects and collects all test cases, infers the testing context from the arguments of polymorphic assertion functions, and selects and executes the appropriate testing protocol. Methods of \texttt{MyTest} whose names begin with \texttt{test} are interpreted as test cases. The \texttt{setUp()} method is used to initialize variables that are shared across all tests.

We introduce two test cases, one for each subroutine. Each test case evaluates the polymorphic assertion function \texttt{self.assertEqual()} with three types of arguments: a distribution function of qubit measurement outcomes, a quantum state represented by a density matrix, and a quantum process represented by a Choi matrix.

After defining the class that represents the testing environment, a test instance is created on line 47 by initializing \texttt{MyTest} with \texttt{AerSimulator} as the backend and specifying 3000 shots to ensure statistical reliability.
The method \texttt{run()} is then called executing the testing protocols. The test outcomes are shown below and correspond to the expected values: as expected, the tests passed for \texttt{subroutine\_correct()} and failed for \texttt{subroutine\_error()} across all types of assertion arguments. The associated probabilities are reported to indicate the confidence levels for the results.

\section{Evaluation}

 \subsection{Usability evaluation}

The results of usability evaluation are reported in Table \ref{tab:evaluation} for unit tests. In the first iteration of the design, the user sets up testing environments by implementing classes that inherit from the child classes of the \texttt{BaseExperiment} interface class from the Qiskit Experiments library (see Figure \ref{fig:uml}). The resulting test script contains up to 82 lines of code and requires 14 imported classes and functions. These imports imply that the user must be familiar with the classes used for quantum tomography, such as \texttt{StateTomography} and \texttt{ProcessTomography}, the fidelity evaluation functions \texttt{state\_fidelity} and \texttt{process\_fidelity}, and the statistical evaluation function \texttt{chisquare} from the \texttt{scipy} library. Additionally, the user must understand the details of the compilation and transpilation processes, particularly the use of the \texttt{transpile} function.

In the second iteration, these implementation details are abstracted away through a unified interface for all testing protocols, supported by a class library of protocol implementations. In this version, the assertion evaluation is handled by a standardized method defined using the same interface for all protocol classes. The user only needs to be aware of the existing protocol classes: \texttt{QUT\_PROJ}, \texttt{QUT\_ST}, and \texttt{QUT\_PT}. As a result, both the number of lines of code and the number of imported entities are reduced by nearly half while accomplishing the same task.

The third iteration of the design introduces an \textit{orchestrator} class, which implements polymorphic assertions and automatically determines the appropriate protocol to invoke based on the type of the arguments (i.e., the context). This has led to further improvements in usability metrics - the task was successfully implemented with just 35 lines of code and only 6 imports. The user does not need to learn the specifications of the classes such as \texttt{QUT\_PROJ}, \texttt{QUT\_ST}, and \texttt{QUT\_PT}. Instead, all test cases are created by inheriting from a single class, \texttt{QUTest} (see Figure \ref{fig:uml}).

Examples of unit test implementations corresponding to each of these design iterations are available at  \url{https://github.com/freude/QUT/tree/main/examples/DSRM_iterations}.

\begin{table}[t]
    \centering
    \caption{Usability metrics}
    \begin{tabular}{|c|c|c|c|}
    \hline
         & DSRM iteration 1 & DSRM iteration 2 & DSRM iteration 3  \\
         \hline
         Lines of codes & 82 & 43 & 35 \\
         Imported entities & 14 & 8 & 6 \\
         \hline
    \end{tabular}
    \label{tab:evaluation}
\end{table}

\subsection{Evaluating unit test quality metrics by mutation analysis}

\begin{figure}
    \centering
    \subfloat[]{\includegraphics[height=0.13\linewidth]{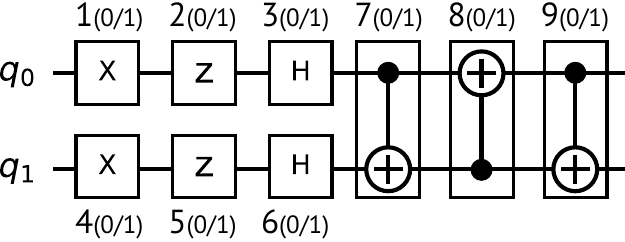}} \hspace{1cm}
    \subfloat[]{\includegraphics[height=0.13\linewidth]{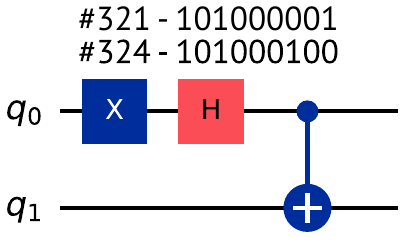}} \hspace{1cm}
     \subfloat[]{\includegraphics[height=0.13\linewidth]{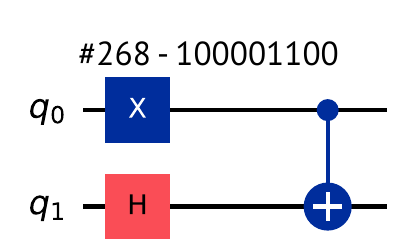}}
    \caption{(a) Template circuit for generating mutation testing samples. The template encodes a nine-bit binary string, where each digit’s position corresponds to a specific quantum gate. A value of 0 indicates the absence, and 1 indicates the presence of the corresponding element. Two example circuits generated from this template are shown: (b) the two-qubit quantum subroutine assumed to be correct, and (c) a variant with a mutated qubit index for the Hadamard gate.}
    \label{fig:circuits}
\end{figure}

To evaluate the quality of results produced by different testing protocols, we apply mutation testing to a small yet representative dataset. The dataset consists of instances of two-qubit circuits constructed from several basic one- and two-qubit gates belonging to the Clifford set. The dataset is generated using the template shown in Figure~\ref{fig:circuits}a. In this template, an integer specifies the position and type of a quantum gate, while a binary value (0 or 1) indicates whether the gate in this position is present in the circuit or absent. If the integer is used as the index of the bit position, each possible circuit diagram - subject to the constraints imposed by the template - can be associated with a nine-bit binary string. This binary string can then be converted into an integer label. The set of mutants is generated by considering all possible combinations of zeros and ones in a nine-bit string, resulting in a total of $2^9 = 512$ circuits.

This method does not guarantee that each circuit is unique; a small number of duplicates may therefore appear in the dataset. For instance, the circuit diagram in Figure~\ref{fig:circuits}b can be represented by the two bit strings $0b101000100$ and $0b101000001$, whose corresponding integer labels are 324 and 321, respectively. The duplicates can be detected and eliminated using quantum process tomography, which provides complete information about a quantum processes in the form of a Choi matrix, Kraus operators, or a Pauli transfer matrix. Nevertheless, a few duplicates, even if retained in the dataset, do not significantly impact the overall results of the mutation testing. Equivalent circuits in the mutant dataset are also commonly observed in other quantum mutation testing frameworks \cite{9678563}. In this work, we retain duplicates in the dataset, assuming that their small number has a statistically insignificant effect on the results. 

The circuit with labels 321 and 324, shown in Figure~\ref{fig:circuits}b, is used as an example of a correct implementation. All other circuits are considered mutants (see example of a mutant in Figure~\ref{fig:circuits}c). The pool of mutants encompasses the following gate-level mutation operators: gate addition, gate deletion, gate replacement, gate retargeting, etc. As an example of gate replacement, the binary strings $0b000000001$ and $0b000000111$ encode the substitution of the CNOT gate with a SWAP gate. Gate retargeting can be exemplified by the pair $0b000000100$ and $0b000000010$, where the gate CNOT(0,1) is replaced by CNOT(1,0).

The experiments are conducted in two configurations: a noise-free setting and a simulated noisy environment. Both configurations were executed using the \texttt{FakeSydneyV2} simulator\footnote{\protect\url{https://docs.quantum.ibm.com/api/qiskit-ibm-runtime/fake-provider-fake-sydney-v2}}, which is part of IBM’s \texttt{Qiskit} class library \cite{qiskit}. This simulator emulates the 27-qubit IBM backend \texttt{ibm\_sydney}. The noise model was derived from historical system snapshots that capture the behavior of the real hardware.

\begin{figure}
    \centering
    \includegraphics[width=\linewidth]{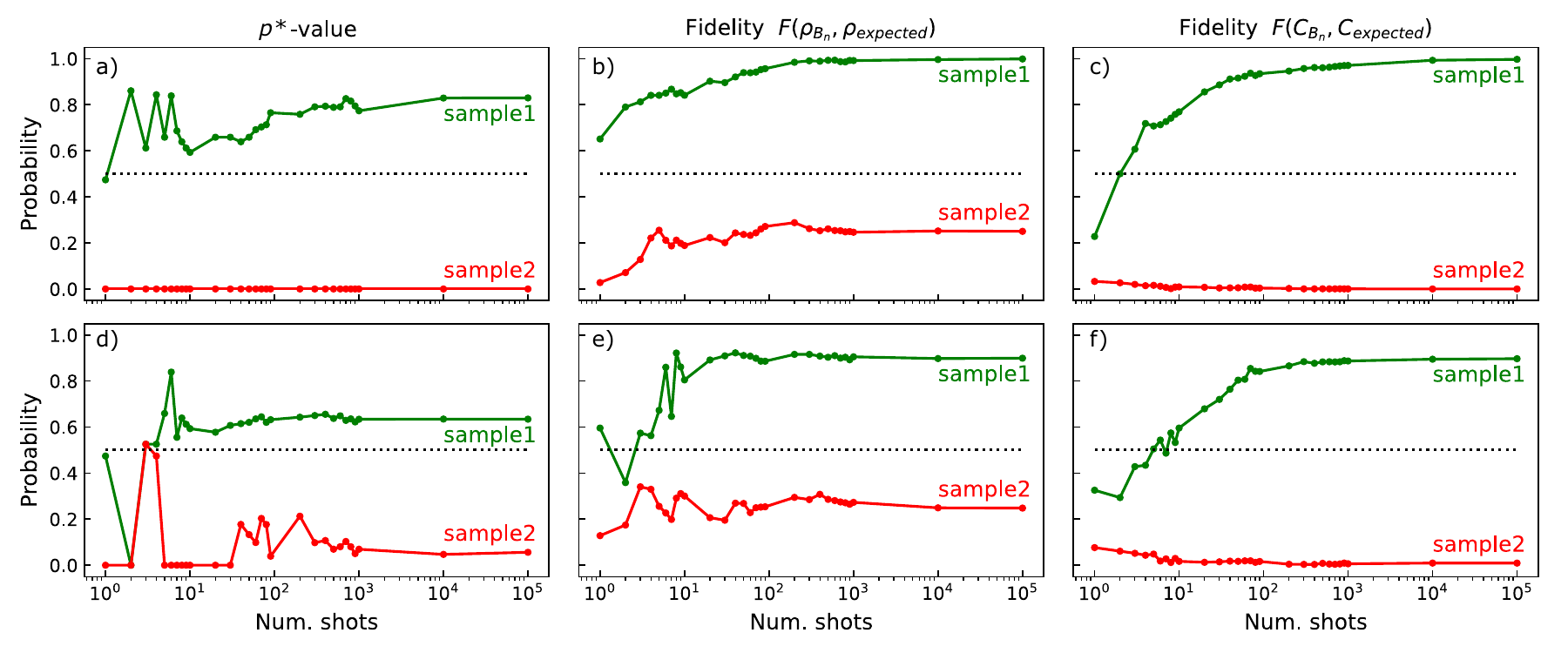}
    \caption{Dependence of the outcomes of the polymorphic probabilistic equality assertion on the number of shots in noiseless environment (panels a, b and c) and with realistic noise (panels d, e and f). The tests have been performed using polymorphic equality assertions with the following arguments: a) and d) the frequency of measurement outcomes, b) and e) density matrices, and c) and f) Choi matrices.}
    \label{fig:results}
\end{figure}

Before performing mutation testing, we first examine how the testing outcomes depend on the number of shots for two quantum subroutines: the correct code (Figure~\ref{fig:circuits}b) and a mutated circuit (Figure~\ref{fig:circuits}c). Figure~\ref{fig:results} shows the probability of a positive test outcome for the correct code (green lines), $\alpha$ in Eq.~(\ref{eq:Youden}), and for the mutated code  (red lines), $\beta$  in Eq.~(\ref{eq:Youden}), across different types of assertions (contexts) - namely, count frequencies of outcomes, quantum states, and quantum processes - applied to the two subroutines defined above.
In all observed cases, $\alpha$ is low for a small number of shots and increases as the number of shots increases. For $\beta$, this dependence is weaker. This implies that, when the number of shots is low, the false negative rate tends to be higher than the false positive rate.

Figure~\ref{fig:results} visualizes $J$-metric as the gap between the green and red curves, representing values of $\alpha$ and $\beta$ respectively, at the highest number of shots. A positive gap (i.e., positive $J$) indicates that the test can successfully detect the error, provided the confidence level is properly set. Conversely, if the gap is closed ($J$ is zero or negative), the test fails to detect the error. It is usually evaluated in the limit of a large number of shots, where the results become independent of the shot count. The statistic takes values in the range -1 to 1, with higher values corresponding to more reliable test outcomes. Negative values reflect an inability to discriminate between correct and incorrect subroutines. 

In noiseless simulations with a large number of shots, the results for the $J$-metric show a high level of confidence in distinguishing between correct and mutated subroutines for process tomography.  In contrast, testing protocols based on the statistical test and state tomography exhibit lower $J$-metric values. The statistical test has a higher probability of false-positive outcomes due to the highly conservative $p$-value calibration: even for large $p$-values, the probability of a positive test outcome $p^*$ remains low. This limitation could be addressed by using a more precise empirical calibration \cite{Wang2026}, although further study of the output data statistics would be required. State tomography, in turn, has a higher probability of false-negative outcomes due to the significant overlap between the expected state and the state produced by the mutated circuit.

The presence of noise reduces the probability of detecting positive cases across all tomographic protocols, while leaving negative-case metrics largely unchanged. For the statistical test, the reliability decreases for both positive and negative test outcomes. In general, noise lowers the $J$-metric for all protocols, reducing the reliability of the tests. 

To extend this analysis beyond a single circuit, we now employ mutation testing to evaluate the accuracy of different testing protocols in detecting bugs across a larger set of circuits. Table~\ref{tab:performance} summarizes the results for the entire dataset obtained with $10^3$ shots for different protocols, with and without hardware noise. In mutation testing, we classify the code as correct when the probability of a positive test outcome exceeds 0.75.  As expected, quantum process tomography successfully identifies all mutants, as the protocol enables the complete reconstruction of the quantum process from the measurement data. State tomography performs slightly worse (96\% of mutants have been detected) while the statistical test performs the worst among the considered approaches (93\% of mutants have been detected). This degradation is not substantial, particularly given the significant differences in computational complexity among these approaches (see Section 6.3). It is important to note that mutants undetected by certain protocols may, in specific contexts, be considered acceptable equivalents of the correct subroutine. This can occur for particular sets of input parameters or specific post-processing procedures. Leveraging such contextual information can substantially reduce computational complexity and should be taken into account when selecting the most appropriate testing protocol. The presence of realistic hardware noise does not significantly affect the number of mutants detected by tomographic protocols, although a small reduction is observed for the protocol based on Pearson’s $\chi^2$ test.

Using Eq.~(\ref{eq:Youden}), we compute the $J$-metric for the entire dataset. For the given set of mutants, process tomography achieves a significantly higher $J$-metric compared to state tomography and Pearson’s $\chi^2$ test. The presence of hardware noise reduces the $J$-metric by approximately 0.1 for both tomographic protocols and by 0.4 for the Pearson's $\chi^2$ test.

We also report $1 - \alpha$ (false-negative rate) and $\beta$ (false-positive rate) metrics. Their analysis allows us to assess whether a probabilistic assertion is biased toward false positives or false negatives and guides the choice of a threshold probability for classifying outcomes. The protocol based on process tomography exhibits small and nearly equal false-positive and false-negative rates, with minimal sensitivity to noise; the presence of hardware noise leads only to a slight increase in the false-negative rate. The protocol based on the quantum state tomography show a pronounced bias toward false-positive outcomes when noise is neglected. Without noise, the Pearson's $\chi^2$ test, like process tomography, exhibits approximately equal false-positive and false-negative rates. Accounting for realistic hardware noise increases the false-negative rate for the tomographic protocols while leaving the false-positive rate unchanged. For the statistical testing protocol, hardware noise also increases the false-negative rate more substantially than the false-positive rate, although its effect on the latter is also non-negligible.

The evaluation results are fully reproducible using the script provided at \url{https://github.com/freude/QUT/tree/main/examples/evaluation.py}.

\begin{table}[t]
    \centering
    \caption{Quality metrics for mutation testing for Pearson's $\chi^2$ test, quantum state tomography, and quantum process tomography, evaluated with and without hardware noise.}
    \begin{tabular}{|p{2.6cm}|>{\centering\arraybackslash}p{2.0cm}|>{\centering\arraybackslash}p{0.7cm}|>{\centering\arraybackslash}p{0.7cm}|>{\centering\arraybackslash}p{0.7cm}|>{\centering\arraybackslash}p{2.0cm}|>{\centering\arraybackslash}p{0.7cm}|>{\centering\arraybackslash}p{0.7cm}|>{\centering\arraybackslash}p{0.7cm}|}
        \hline
        &\multicolumn{4}{|c|}{Noiseless} & \multicolumn{4}{|c|}{With noise}   \\
    \cline{2-9}
                 Testing protocol & Fraction of mutants killed& $J$ & $1-\alpha$ (FNR) & $\beta$ (FPR) & Fraction of mutants killed& $J$ &  $1-\alpha$ (FNR) & $\beta$ (FPR)  \\
         \hline
         Pearson's $\chi^2$-test &  0.93 &  0.58 &  0.23 & 0.19 & 0.67 & 0.40 & 0.37 & 0.23\\
         State tomography          &  0.96 &  0.77 & 0.01  & 0.22 & 0.96 & 0.68 & 0.10 & 0.22\\
         Process tomography     &  1.00 &  0.91 & 0.03  & 0.06  & 1.00 &  0.83 & 0.11 & 0.06\\
         \hline
    \end{tabular}
    \label{tab:performance}
\end{table}

\subsection{Performance and scalability}

We have also measured the average execution time per shot (i.e., wall time divided by the number of shots) for each type of assertion argument, with all parallelization techniques disabled. The execution time per shot is $5.47 \times 10^{-7}$ s for the distribution function equality assertion (Pearson’s $\chi^2$ test), $1.22 \times 10^{-5}$ s for the density matrix equality assertion (quantum state tomography), and $2.02 \times 10^{-4}$ s for the Choi matrix equality  assertion (quantum process tomography).

\begin{figure}
    \centering
    \includegraphics[width=0.8\linewidth]{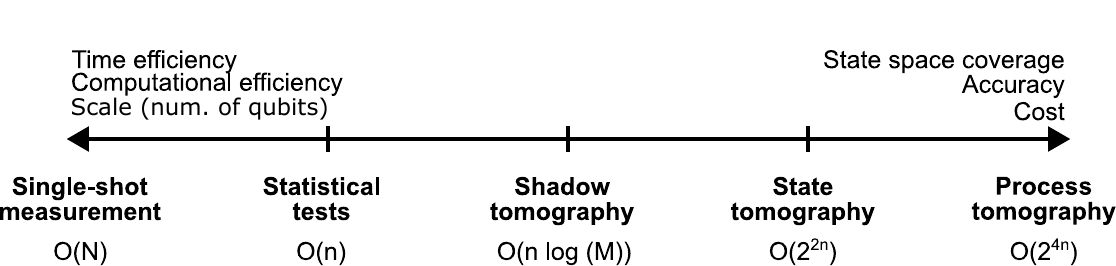}
    \caption{Qualitative comparison of quantum testing protocols, highlighting the trade-offs between computational efficiency, state-space coverage, accuracy, cost, and the scale of problems that can be tackled. The protocols include single-shot measurement, statistical tests, shadow tomography, state tomography, and process tomography, each offering different trade-offs between testing cost and the amount of information obtained about the quantum system.  The diagram also shows the asymptotic computational complexities of the testing protocols using Big-O-notation, where $N$ is a number of qubits and $n=N \cdot \alpha$, where $\alpha$ is a number of shots, and $M$ is the number of observables.}
    \label{fig:scalability}
\end{figure}

Protocols relying on statistical tests or quantum tomography have established computational complexity that grows with the number of qubits, directly impacting the scalability of the entire testing process. Based on results reported in the literature \cite{PhysRevA.77.032322, paris2004quantum, Huang2019, Huang2020}, we summarise the computational complexity of different testing protocols in the diagram in Figure~\ref{fig:scalability}, together with the associated trade-offs.

Protocols based on single-shot measurements and quantum process tomography represent two extreme cases and are impractical in most situations: the former is generally uninformative, while the latter is applicable only to circuits with a very small number of qubits due to its exponential complexity. Testing protocols occupying the middle of the scale in Figure~\ref{fig:scalability} are therefore particularly relevant in practice. The complexity of statistical tests scales linearly with the number of qubits, whereas the complexity of state tomography grows exponentially, although still more slowly than that of process tomography. Quantum shadow tomography, which we did not consider in this work, can also be regarded as a form of the protocol based on statistical tests and may represent a promising candidate for unit testing protocols when assertions involve expectation values of observables. 

The wall times of tests and accuracy observed during the first design iteration remained consistent across all subsequent iterations within the DSRM. These metrics are determined by a specific testing protocol, rather than by the design details themselves. The design influences performance and accuracy only through the selection of particular testing protocols that have been incorporated into the protocol library, where they become an integral part of the framework.

\subsection{Answers to research questions}

\begin{enumerate}[label={\textit{RQ\arabic*}}]
\item The testing protocols differ in their coverage of the quantum input and output state spaces. The protocol based on process tomography provides complete information about the quantum process underlying the denotational semantics of a quantum subroutine. However, due to its exponential computational complexity, it is feasible in practice only for subroutines with a very small number of qubits. State tomography and statistical tests are less computationally demanding, but they do not guarantee 100\% bug detection and must therefore be applied with caution. The choice of testing protocol can be guided by contextual information indicating which parts of the state spaces covered by the subroutine will be used in the application. The quantum state tomography protocol exhibits a higher false-positive rate than process tomography due to uncertainty arising from the potential overlap between the expected state and the state produced by faulty code. However, the false-positive rate is expected to decrease for larger systems, as the increased dimensionality of the Hilbert space makes accidental state overlap less probable. The testing protocol based on the $\chi^2$ statistical test is even less reliable, primarily because the output $p$-value cannot be directly interpreted as the probability of a Type I error and therefore requires a calibration procedure. The calibration procedure tends to overestimate error probabilities when no additional information about the data is available.

\item Hardware noise reduces the overall ability of unit tests to discriminate between correct and mutated subroutines. This reduction corresponds to approximately 0.1–0.2 in the $J$-metric. Hardware noise also increases the false-negative rate, while having a smaller effect on the false-positive rate. Recognizing that noise can impair the precision of quantum unit testing results even in the limit of infinitely many shots, we suggest applying quantum error correction \cite{Roffe03072019} and/or error mitigation \cite{RevModPhys.95.045005} techniques along with unit testing procedures. The testing protocol based on classical statistical tests is the most sensitive to noise, as noise can substantially alter the statistics of the output data, increasing both false-positive and false-negative rates.

\item Contextual information can reduce the dimensionality of the input parameter space for subroutines, as well as the dimensionality of the relevant quantum information produced by a subroutine. Generally, the denotational semantics of a quantum subroutine is represented by a quantum process. The context determines whether the application uses the full information encoded in this process or only a part of it, such as a particular output state, a projection of a state, or a deterministic or random variable associated with a specific observable. This reduction changes the semantics of the subroutine and suggests alternative arguments for assertion statements. This makes the use of process tomography unnecessary in many specific cases and motivates the adoption of alternative testing protocols with lower coverage but reduced computational complexity. The proposed framework adapts to the assertion variables selected by the user by choosing the appropriate testing protocol.
\end{enumerate}

\subsection{Future work}

The current framework derives contextual information from user-provided argument types rather than analyzing the source code automatically. Future work will explore inferring contextual information from the semantics and structure of quantum programs or user-provided project descriptions to improve the automation of context-aware test generation.

The current prototype supports three testing protocols, with shadow tomography and other protocols left for future integration. The framework is designed to be easily extended with a larger library of testing protocols.

\section{Threats to Validity}

The validity of the usability evaluation may be influenced by somewhat subjective choices in complexity measurement, such as using the number of lines of code or imported entities. Although these are reasonable indicators, they may not fully capture the cognitive load or learning curve experienced by users. Furthermore, no user studies or empirical testing with real developers were conducted to verify whether the observed improvements translate into a better user experience in practice. In mutation testing, internal validity may also be affected by the assumption that the selected dataset - consisting of two-qubit subroutines with one- and two-qubit gates - is sufficiently representative to capture the most common code mutations arising from hardware and human errors. 

The experimental evaluation is conducted exclusively on quantum simulators, including simulations of quantum noise. Although these simulations enable controlled evaluation under configurable noise models, validation on physical quantum hardware remains a direction for future work to assess the framework under actual device constraints and noise characteristics.

\section{Conclusion}

We introduced QUT, a novel unit testing framework for quantum subroutines that combines architectural design principles with a practical prototype implementation. QUT provides a unified interface for probabilistic context-dependent assertions, automatically selecting suitable testing protocols -- such as Pearson’s chi-squared test, quantum state tomography, or quantum process tomography -- based on the semantics of the subroutine under test. Built on the Qiskit stack, the framework ensures compatibility with both simulation environments and real IBM quantum hardware.

The evaluation demonstrated that QUT improves usability by substantially reducing the amount of code and prior knowledge required from users, while maintaining accuracy in detecting faulty subroutines. The user is required to provide only arguments of a specific type in the assertions, while the framework automatically selects the appropriate testing protocol, taking into account the context-dependent semantics of quantum subroutines. The experimental results confirm that the framework can reliably distinguish between correct and erroneous implementations under noiseless conditions, as well as in the noisy environment inherent to current NISQ devices, albeit with slightly reduced accuracy.  These findings underscore the importance of context-aware polymorphic assertions and automated protocol orchestration in making quantum unit testing practical.

Beyond its immediate contributions, QUT represents a step toward integrating rigorous testing practices into the quantum software development lifecycle, in alignment with DevOps-inspired methodologies for hybrid quantum-classical systems. By lowering the barrier to writing effective quantum unit tests, the framework facilitates more reliable and maintainable software development as quantum programs grow in size and complexity.

Future work will extend QUT in several directions. First, expanding the protocol library with additional statistical and tomographic techniques will enhance its coverage and adaptability. Second, user studies and integration into CI/CD pipelines will provide deeper insights into its usability and adoption in real-world quantum software projects. Finally, exploring error mitigation strategies and backend-specific optimizations will be crucial for improving confidence levels under realistic noisy hardware conditions.

\bibliographystyle{unsrt}
\bibliography{bib}

@inproceedings{Huang2019,
author = {Huang, Yipeng and Martonosi, Margaret},
title = {Statistical assertions for validating patterns and finding bugs in quantum programs},
year = {2019},
isbn = {9781450366694},
publisher = {Association for Computing Machinery},
address = {New York, NY, USA},
url = {https://doi.org/10.1145/3307650.3322213},
doi = {10.1145/3307650.3322213},
booktitle = {Proceedings of the 46th International Symposium on Computer Architecture},
pages = {541–553},
numpages = {13},
location = {Phoenix, Arizona},
series = {ISCA '19}
}

@Article{Huang2020,
author={Huang, Hsin-Yuan
and Kueng, Richard
and Preskill, John},
title={Predicting many properties of a quantum system from very few measurements},
journal={Nature Physics},
year={2020},
month={Oct},
day={01},
volume={16},
number={10},
pages={1050-1057},
issn={1745-2481},
doi={10.1038/s41567-020-0932-7},
url={https://doi.org/10.1038/s41567-020-0932-7}
}

@article{PhysRevLett.86.4195,
  title = {Quantum Tomography for Measuring Experimentally the Matrix Elements of an Arbitrary Quantum Operation},
  author = {D'Ariano, G. M. and Lo Presti, P.},
  journal = {Phys. Rev. Lett.},
  volume = {86},
  issue = {19},
  pages = {4195--4198},
  numpages = {0},
  year = {2001},
  month = {May},
  publisher = {American Physical Society},
  doi = {10.1103/PhysRevLett.86.4195},
  url = {https://link.aps.org/doi/10.1103/PhysRevLett.86.4195}
}

@book{Nielsen_Chuang_2010, place={Cambridge}, title={Quantum Computation and Quantum Information: 10th Anniversary Edition}, publisher={Cambridge University Press}, author={Nielsen, Michael A. and Chuang, Isaac L.}, year={2010}}

@book{Wilde_2013, 
place={Cambridge}, 
title={Quantum Information Theory},
publisher={Cambridge University Press}, 
author={Wilde, Mark M.},
year={2013}}

@article{Audenaert,
  title = {Discriminating States: The Quantum Chernoff Bound},
  author = {Audenaert, K. M. R. and Calsamiglia, J. and Mu\~noz-Tapia, R. and Bagan, E. and Masanes, Ll. and Acin, A. and Verstraete, F.},
  journal = {Phys. Rev. Lett.},
  volume = {98},
  issue = {16},
  pages = {160501},
  numpages = {4},
  year = {2007},
  month = {Apr},
  publisher = {American Physical Society},
  doi = {10.1103/PhysRevLett.98.160501},
  url = {https://link.aps.org/doi/10.1103/PhysRevLett.98.160501}
}

@book{tomamichel2015quantum,
  title={Quantum Information Processing with Finite Resources: Mathematical Foundations},
  author={Tomamichel, M.},
  isbn={9783319218915},
  series={SpringerBriefs in Mathematical Physics},
  url={https://books.google.com.au/books?id=643DCgAAQBAJ},
  year={2015},
  publisher={Springer International Publishing}
}

@article{PhysRevA.64.052312,
  title = {Measurement of qubits},
  author = {James, Daniel F. V. and Kwiat, Paul G. and Munro, William J. and White, Andrew G.},
  journal = {Phys. Rev. A},
  volume = {64},
  issue = {5},
  pages = {052312},
  numpages = {15},
  year = {2001},
  month = {Oct},
  publisher = {American Physical Society},
  doi = {10.1103/PhysRevA.64.052312},
  url = {https://link.aps.org/doi/10.1103/PhysRevA.64.052312}
}

@article{Renato_2016,
  title = {Practical and Reliable Error Bars in Quantum Tomography},
  author = {Faist, Philippe and Renner, Renato},
  journal = {Phys. Rev. Lett.},
  volume = {117},
  issue = {1},
  pages = {010404},
  numpages = {6},
  year = {2016},
  month = {Jul},
  publisher = {American Physical Society},
  doi = {10.1103/PhysRevLett.117.010404},
  url = {https://link.aps.org/doi/10.1103/PhysRevLett.117.010404}
}

@article{Smolin_2012,
  title = {Efficient Method for Computing the Maximum-Likelihood Quantum State from Measurements with Additive Gaussian Noise},
  author = {Smolin, John A. and Gambetta, Jay M. and Smith, Graeme},
  journal = {Phys. Rev. Lett.},
  volume = {108},
  issue = {7},
  pages = {070502},
  numpages = {4},
  year = {2012},
  month = {Feb},
  publisher = {American Physical Society},
  doi = {10.1103/PhysRevLett.108.070502},
  url = {https://link.aps.org/doi/10.1103/PhysRevLett.108.070502}
}

@article{Renner_2012,
  title = {Reliable Quantum State Tomography},
  author = {Christandl, Matthias and Renner, Renato},
  journal = {Phys. Rev. Lett.},
  volume = {109},
  issue = {12},
  pages = {120403},
  numpages = {6},
  year = {2012},
  month = {Sep},
  publisher = {American Physical Society},
  doi = {10.1103/PhysRevLett.109.120403},
  url = {https://link.aps.org/doi/10.1103/PhysRevLett.109.120403}
}

@article{Renner_2019,
  title = {Confidence Polytopes in Quantum State Tomography},
  author = {Wang, Jinzhao and Scholz, Volkher B. and Renner, Renato},
  journal = {Phys. Rev. Lett.},
  volume = {122},
  issue = {19},
  pages = {190401},
  numpages = {6},
  year = {2019},
  month = {May},
  publisher = {American Physical Society},
  doi = {10.1103/PhysRevLett.122.190401},
  url = {https://link.aps.org/doi/10.1103/PhysRevLett.122.190401}
}

@misc{qiskit,
      title={{Quantum computing with Qiskit}}, 
      author={Ali Javadi-Abhari and Matthew Treinish and Kevin Krsulich and Christopher J. Wood and Jake Lishman and Julien Gacon and Simon Martiel and Paul D. Nation and Lev S. Bishop and Andrew W. Cross and Blake R. Johnson and Jay M. Gambetta},
      year={2024},
      eprint={2405.08810},
      archivePrefix={arXiv},
      primaryClass={quant-ph},
      url={https://arxiv.org/abs/2405.08810}, 
}

@article{chuang1997prescription,
  title={Prescription for experimental determination of the dynamics of a quantum black box},
  author={Chuang, Isaac L and Nielsen, Michael A},
  journal={Journal of Modern Optics},
  volume={44},
  number={11-12},
  pages={2455--2467},
  year={1997},
  publisher={Taylor \& Francis}
}

@inproceedings{10.1145/3188745.3188802,
author = {Aaronson, Scott},
title = {Shadow tomography of quantum states},
year = {2018},
isbn = {9781450355599},
publisher = {Association for Computing Machinery},
address = {New York, NY, USA},
url = {https://doi.org/10.1145/3188745.3188802},
doi = {10.1145/3188745.3188802},
booktitle = {Proceedings of the 50th Annual ACM SIGACT Symposium on Theory of Computing},
pages = {325–338},
numpages = {14},
location = {Los Angeles, CA, USA},
series = {STOC 2018}
}

@article{PhysRevA.77.032322,
  title = {Quantum-process tomography: Resource analysis of different strategies},
  author = {Mohseni, M. and Rezakhani, A. T. and Lidar, D. A.},
  journal = {Phys. Rev. A},
  volume = {77},
  issue = {3},
  pages = {032322},
  numpages = {15},
  year = {2008},
  month = {Mar},
  publisher = {American Physical Society},
  doi = {10.1103/PhysRevA.77.032322},
  url = {https://link.aps.org/doi/10.1103/PhysRevA.77.032322}
}

@article{qasm3,
author = {Cross, Andrew and Javadi-Abhari, Ali and Alexander, Thomas and De Beaudrap, Niel and Bishop, Lev S. and Heidel, Steven and Ryan, Colm A. and Sivarajah, Prasahnt and Smolin, John and Gambetta, Jay M. and Johnson, Blake R.},
title = {OpenQASM 3: A Broader and Deeper Quantum Assembly Language},
year = {2022},
issue_date = {September 2022},
publisher = {Association for Computing Machinery},
address = {New York, NY, USA},
volume = {3},
number = {3},
url = {https://doi.org/10.1145/3505636},
doi = {10.1145/3505636},
journal = {ACM Transactions on Quantum Computing},
month = sep,
articleno = {12},
numpages = {50}
}

@article{PhysRevA.93.042316,
  title = {Generalized Hofmann quantum process fidelity bounds for quantum filters},
  author = {Sedl\'ak, Michal and Fiur\'a\ifmmode \check{s}\else \v{s}\fi{}ek, Jarom\'{\i}r},
  journal = {Phys. Rev. A},
  volume = {93},
  issue = {4},
  pages = {042316},
  numpages = {10},
  year = {2016},
  month = {Apr},
  publisher = {American Physical Society},
  doi = {10.1103/PhysRevA.93.042316},
  url = {https://link.aps.org/doi/10.1103/PhysRevA.93.042316}
}

@book{bass2003software,
  title={Software Architecture in Practice},
  author={Bass, L. and Clements, P. and Kazman, R.},
  isbn={9780321154958},
  lccn={2003045300},
  series={SEI series in software engineering},
  url={https://books.google.com.au/books?id=mdiIu8Kk1WMC},
  year={2003},
  publisher={Addison-Wesley}
}

@ARTICLE{10463159,
  author={Metwalli, Sara Ayman and Van Meter, Rodney},
  journal={IEEE Transactions on Quantum Engineering}, 
  title={Testing and Debugging Quantum Circuits}, 
  year={2024},
  volume={5},
  number={},
  pages={1-15},
  doi={10.1109/TQE.2024.3374879}}

@article{10.1145/3656339,
author = {Long, Peixun and Zhao, Jianjun},
title = {Testing Multi-Subroutine Quantum Programs: From Unit Testing to Integration Testing},
year = {2024},
issue_date = {July 2024},
publisher = {Association for Computing Machinery},
address = {New York, NY, USA},
volume = {33},
number = {6},
issn = {1049-331X},
url = {https://doi.org/10.1145/3656339},
doi = {10.1145/3656339},
journal = {ACM Trans. Softw. Eng. Methodol.},
month = jun,
articleno = {147},
numpages = {61}
}

@article{klymenko2024architectural,
title = {Architectural patterns for designing quantum artificial intelligence systems},
journal = {Journal of Systems and Software},
volume = {227},
pages = {112456},
year = {2025},
issn = {0164-1212},
doi = {https://doi.org/10.1016/j.jss.2025.112456},
url = {https://www.sciencedirect.com/science/article/pii/S0164121225001244},
author = {Mykhailo Klymenko and Thong Hoang and Xiwei Xu and Zhenchang Xing and Muhammad Usman and Qinghua Lu and Liming Zhu}
}

@article{garcia2023quantum,
  title={Quantum software testing: State of the art},
  author={Garc{\'\i}a de la Barrera, Antonio and Garc{\'\i}a-Rodr{\'\i}guez de Guzm{\'a}n, Ignacio and Polo, Macario and Piattini, Mario},
  journal={Journal of Software: Evolution and Process},
  volume={35},
  number={4},
  pages={e2419},
  year={2023},
  publisher={Wiley Online Library}
}

@book{vaughan2013scientific,
  title={Scientific Inference: Learning from Data},
  author={Vaughan, S.},
  isbn={9781107024823},
  lccn={2013021427},
  series={Scientific Inference: Learning from Data},
  url={https://books.google.com.au/books?id=HdWaAAAAQBAJ},
  year={2013},
  publisher={Cambridge University Press}
}

@book{gamma1995design,
  title={Design Patterns: Elements of Reusable Object-Oriented Software},
  author={Gamma, E.},
  isbn={9788131700075},
  series={Addison-Wesley professional computing series},
  url={https://books.google.com.au/books?id=K4qv1D-LKhoC},
  year={1995},
  publisher={Pearson Education}
}

@article{klymenko2025,
doi = {10.1088/2058-9565/ae629e},
url = {https://doi.org/10.1088/2058-9565/ae629e},
year = {2026},
month = {apr},
publisher = {IOP Publishing},
volume = {11},
number = {2},
pages = {025045},
author = {Klymenko, Mykhailo and Hoang, Thong and Wilkinson, Samuel A and Goldozian, Bahar and Ma, Suyu and Xu, Xiwei and Lu, Qinghua and Usman, Muhammad and Zhu, Liming},
title = {Context-aware unit testing for quantum subroutines},
journal = {Quantum Science and Technology},
}

@article{Peffers01122007,
author = {Ken Peffers and Tuure Tuunanen and Marcus A. Rothenberger and Samir Chatterjee},
title = {A Design Science Research Methodology for Information Systems Research},
journal = {Journal of Management Information Systems},
volume = {24},
number = {3},
pages = {45--77},
year = {2007},
publisher = {Routledge},
doi = {10.2753/MIS0742-1222240302},
}

@book{Heunen_book,
    author = {Heunen, Chris and Vicary, Jamie},
    title = {Categories for Quantum Theory: An Introduction},
    publisher = {Oxford University Press},
    year = {2019},
    month = {11},
    isbn = {9780198739623},
    doi = {10.1093/oso/9780198739623.001.0001},
}

@book{gay2010semantic,
  title={Semantic Techniques in Quantum Computation},
  author={Gay, S. and Mackie, I.},
  isbn={9780521513746},
  lccn={2009032804},
  url={https://books.google.com.au/books?id=__LnWv6NYM0C},
  year={2010},
  publisher={Cambridge University Press}
}

@book{coecke2017picturing,
  title={Picturing Quantum Processes: A First Course in Quantum Theory and Diagrammatic Reasoning},
  author={Coecke, B. and Kissinger, A.},
  isbn={9781108107716},
  url={https://books.google.com.au/books?id=Tn47DgAAQBAJ},
  year={2017},
  publisher={Cambridge University Press}
}

@article{SELINGER2011113,
title = {Finite Dimensional Hilbert Spaces are Complete for Dagger Compact Closed Categories (Extended Abstract)},
journal = {Electronic Notes in Theoretical Computer Science},
volume = {270},
number = {1},
pages = {113-119},
year = {2011},
note = {Proceedings of the Joint 5th International Workshop on Quantum Physics and Logic and 4th Workshop on Developments in Computational Models (QPL/DCM 2008)},
issn = {1571-0661},
doi = {https://doi.org/10.1016/j.entcs.2011.01.010}
}

@misc{miranskyy2025,
      title={On the Feasibility of Quantum Unit Testing}, 
      author={Andriy Miranskyy and José Campos and Anila Mjeda and Lei Zhang and Ignacio García Rodríguez de Guzmán},
      year={2025},
      eprint={2507.17235},
      archivePrefix={arXiv},
      primaryClass={cs.SE},
      url={https://arxiv.org/abs/2507.17235}, 
}

@ARTICLE{10859187,
  author={Wei, Chenhao and Xiao, Lu and Yu, Tingting and Wong, Sunny and Clune, Abigail},
  journal={IEEE Transactions on Software Engineering}, 
  title={How Do Developers Structure Unit Test Cases? An Empirical Analysis of the AAA Pattern in Open Source Projects}, 
  year={2025},
  volume={51},
  number={4},
  pages={1007-1038},
  doi={10.1109/TSE.2025.3537337}}

@InProceedings{10.1007/978-3-662-54434-1_14,
author="Culpepper, Ryan
and Cobb, Andrew",
editor="Yang, Hongseok",
title="Contextual Equivalence for Probabilistic Programs with Continuous Random Variables and Scoring",
booktitle="Programming Languages and Systems",
year="2017",
publisher="Springer Berlin Heidelberg",
address="Berlin, Heidelberg",
pages="368--392",
isbn="978-3-662-54434-1"
}

@InProceedings{10.1007/978-3-540-71316-6_1,
author="Pitts, Andrew",
editor="De Nicola, Rocco",
title="Techniques for Contextual Equivalence in Higher-Order, Typed Languages",
booktitle="Programming Languages and Systems",
year="2007",
publisher="Springer Berlin Heidelberg",
address="Berlin, Heidelberg",
pages="1--1",
isbn="978-3-540-71316-6"
}

@article{UHLMANN1976273,
title = {{The “transition probability” in the state space of a *-algebra}},
journal = {Reports on Mathematical Physics},
volume = {9},
number = {2},
pages = {273-279},
year = {1976},
issn = {0034-4877},
doi = {https://doi.org/10.1016/0034-4877(76)90060-4},
url = {https://www.sciencedirect.com/science/article/pii/0034487776900604},
author = {A. Uhlmann}
}

@article{Jozsa01121994,
author = {Richard Jozsa},
title = {Fidelity for Mixed Quantum States},
journal = {Journal of Modern Optics},
volume = {41},
number = {12},
pages = {2315--2323},
year = {1994},
publisher = {Taylor \& Francis},
doi = {10.1080/09500349414552171}
}

@misc{python-unittest,
  author       = {{Python Software Foundation}},
  title        = {{unittest — Unit testing framework}},
  year         = {2024},
  howpublished = {\url{https://docs.python.org/3/library/unittest.html}},
  note         = {Accessed: 2025-07-31}
}

@misc{qiskit-experiments,
  author       = {Qiskit Community},
  title        = {Qiskit Experiments},
  year         = {2024},
  url          = {https://qiskit-community.github.io/qiskit-experiments/},
  note         = {Accessed: 2025-08-03},
  howpublished = {\url{https://qiskit-community.github.io/qiskit-experiments/}}
}

@article{Preskill2018quantumcomputingin,
  doi = {10.22331/q-2018-08-06-79},
  url = {https://doi.org/10.22331/q-2018-08-06-79},
  title = {Quantum {C}omputing in the {NISQ} era and beyond},
  author = {Preskill, John},
  journal = {{Quantum}},
  issn = {2521-327X},
  publisher = {{Verein zur F{\"{o}}rderung des Open Access Publizierens in den Quantenwissenschaften}},
  volume = {2},
  pages = {79},
  month = aug,
  year = {2018}
}

@Article{Carroll2022,
author={Carroll, M.
and Rosenblatt, S.
and Jurcevic, P.
and Lauer, I.
and Kandala, A.},
title={Dynamics of superconducting qubit relaxation times},
journal={npj Quantum Information},
year={2022},
month={Nov},
day={17},
volume={8},
number={1},
pages={132},
issn={2056-6387},
doi={10.1038/s41534-022-00643-y}
}

@inproceedings{10.1109/ASE51524.2021.9678792,
author = {Wang, Jiyuan and Zhang, Qian and Xu, Guoqing Harry and Kim, Miryung},
title = {QDiff: differential testing of quantum software stacks},
year = {2022},
isbn = {9781665403375},
publisher = {IEEE Press},
url = {https://doi.org/10.1109/ASE51524.2021.9678792},
doi = {10.1109/ASE51524.2021.9678792},
booktitle = {Proceedings of the 36th IEEE/ACM International Conference on Automated Software Engineering},
pages = {692–704},
numpages = {13},
location = {Melbourne, Australia},
series = {ASE '21}
}

@INPROCEEDINGS{9678563,
  author={Mendiluze, Eñaut and Ali, Shaukat and Arcaini, Paolo and Yue, Tao},
  booktitle={2021 36th IEEE/ACM International Conference on Automated Software Engineering (ASE)}, 
  title={Muskit: A Mutation Analysis Tool for Quantum Software Testing}, 
  year={2021},
  volume={},
  number={},
  pages={1266-1270},
  doi={10.1109/ASE51524.2021.9678563}}

@ARTICLE{9844849,
  author={Fortunato, Daniel and CAMPOS, JOSÉ and ABREU, RUI},
  journal={IEEE Transactions on Quantum Engineering}, 
  title={Mutation Testing of Quantum Programs: A Case Study With Qiskit}, 
  year={2022},
  volume={3},
  number={},
  pages={1-17},
  doi={10.1109/TQE.2022.3195061}}

@INPROCEEDINGS{9678798,
  author={Wang, Xinyi and Arcaini, Paolo and Yue, Tao and Ali, Shaukat},
  booktitle={2021 36th IEEE/ACM International Conference on Automated Software Engineering (ASE)}, 
  title={Quito: a Coverage-Guided Test Generator for Quantum Programs}, 
  year={2021},
  volume={},
  number={},
  pages={1237-1241},
  doi={10.1109/ASE51524.2021.9678798}}

@article{quckeck,
author = {{Pontolillo, Gabriel and Mousavi, Mohammad Reza and Grzesiuk, Marek}},
title = {{QuCheck: A Property-based Testing Framework for Quantum Programs in Qiskit}},
year = {2026},
issue_date = {December 2026},
publisher = {Association for Computing Machinery},
address = {New York, NY, USA},
volume = {7},
number = {4},
url = {https://doi.org/10.1145/3815169},
doi = {10.1145/3815169},
journal = {ACM Transactions on Quantum Computing},
month = aug,
articleno = {24},
numpages = {32}
}

@book{gordon2012denotational,
  title={The Denotational Description of Programming Languages: An Introduction},
  author={Gordon, M.J.C.},
  isbn={9781461262282},
  lccn={79015723},
  series={Computer Science},
  url={https://books.google.com.au/books?id=s4jTBwAAQBAJ},
  year={2012},
  publisher={Springer New York}
}

@article{KLYMENKO2021107676,
title = {{NanoNET: An extendable Python framework for semi-empirical tight-binding models}},
journal = {Computer Physics Communications},
volume = {259},
pages = {107676},
year = {2021},
issn = {0010-4655},
doi = {https://doi.org/10.1016/j.cpc.2020.107676},
url = {https://www.sciencedirect.com/science/article/pii/S0010465520303283},
author = {M.V. Klymenko and J.A. Vaitkus and J.S. Smith and J.H. Cole}
}

@book{lutz2009learning,
  title={Learning Python: Powerful Object-Oriented Programming},
  author={Lutz, M.},
  isbn={9781449379322},
  lccn={2010485706},
  series={Animal Guide},
  url={https://books.google.com.au/books?id=1HxWGezDZcgC},
  year={2009},
  publisher={O'Reilly Media}
}

@article{miranskyy2025feasibility,
  title={On the Feasibility of Quantum Unit Testing},
  author={Miranskyy, Andriy and Campos, Jos{\'e} and Mjeda, Anila and Zhang, Lei and de Guzman, Ignacio Garcia Rodriguez},
  journal={arXiv preprint arXiv:2507.17235},
  year={2025}
}

@book{paris2004quantum,
  title={Quantum State Estimation},
  author={Paris, M. and Rehacek, J.},
  isbn={9783540223290},
  lccn={2004107804},
  series={Lecture Notes in Physics},
  url={https://books.google.com.au/books?id=Grr25VFtGgUC},
  year={2004},
  publisher={Springer Berlin Heidelberg}
}

@article{ABBADANDALOUSSI2023111619,
title = {On the relationship between source-code metrics and cognitive load: A systematic tertiary review},
journal = {Journal of Systems and Software},
volume = {198},
pages = {111619},
year = {2023},
issn = {0164-1212},
doi = {https://doi.org/10.1016/j.jss.2023.111619},
url = {https://www.sciencedirect.com/science/article/pii/S0164121223000146},
author = {Amine Abbad-Andaloussi},
}

@article{sellke2001calibration,
  title={Calibration of $\rho$ values for testing precise null hypotheses},
  author={Sellke, Thomas and Bayarri, Mar{\'\i}a Jes{\'u}s and Berger, James O},
  journal={The American Statistician},
  volume={55},
  number={1},
  pages={62--71},
  year={2001},
  publisher={Taylor \& Francis}
}

@article{10.1145/3838727,
author = {Zappin, Jake and Stalnaker, Trevor and Chaparro, Oscar and Poshyvanyk, Denys},
title = {Challenges and Practices in Quantum Software Testing and Debugging: Insights from Practitioners},
year = {2026},
publisher = {Association for Computing Machinery},
address = {New York, NY, USA},
issn = {1049-331X},
url = {https://doi.org/10.1145/3838727},
doi = {10.1145/3838727},
journal = {ACM Trans. Softw. Eng. Methodol.},
month = aug
}

@article{RevModPhys.95.045005,
  title = {Quantum error mitigation},
  author = {Cai, Zhenyu and Babbush, Ryan and Benjamin, Simon C. and Endo, Suguru and Huggins, William J. and Li, Ying and McClean, Jarrod R. and O'Brien, Thomas E.},
  journal = {Rev. Mod. Phys.},
  volume = {95},
  issue = {4},
  pages = {045005},
  numpages = {37},
  year = {2023},
  month = {Dec},
  publisher = {American Physical Society},
  doi = {10.1103/RevModPhys.95.045005},
  url = {https://link.aps.org/doi/10.1103/RevModPhys.95.045005}
}

@article{Roffe03072019,
author = {Joschka Roffe},
title = {Quantum error correction: an introductory guide},
journal = {Contemporary Physics},
volume = {60},
number = {3},
pages = {226--245},
year = {2019},
publisher = {Taylor \& Francis},
doi = {10.1080/00107514.2019.1667078},
}

@Article{Wang2026,
author={Wang, Mengyao
and Li, Wenwen
and Lu, Zhenzhen
and Wang, Yingying
and Huang, Lihong
and Chen, Feng},
title={Evaluating empirical calibration of <em>P</em> values under unmeasured confounding bias: a simulation study and real-world application},
journal={Journal of Clinical Epidemiology},
year={2026},
month={May},
day={01},
publisher={Elsevier},
volume={193},
issn={0895-4356},
doi={10.1016/j.jclinepi.2026.112188},
url={https://doi.org/10.1016/j.jclinepi.2026.112188}
}

\end{document}